# The ro-vibrational $v_2$ mode spectrum of methane investigated by ultrabroadband coherent Raman spectroscopy

**Francesco Mazza,[1,a)] Ona Thornquist,[1] Leonardo Castellanos,[1] Thomas Butterworth,[2] Cyril Richard,[3] Vincent Boudon,[3] and Alexis Bohlin[1,4]**

[1]Faculty of Aerospace Engineering, Delft University of Technology, Kluyverweg 1, 2629 HS Delft, The Netherlands

[2]Faculty of Science and Engineering, Maastricht University, Paul Henri Spaaklaan 1, 6229 GS Maastricht, The Netherlands

[3]Laboratoire Interdisciplinaire Carnot de Bourgogne, UMR 6303 CNRS - Université Bourgogne Franche-Comté 9 Avenue Alain Savary, BP 47 870, F-21078 Dijon Cedex, France

[4]Space Propulsion Laboratory, Department of Computer Science, Electrical and Space Engineering, Luleå University of Technology, Bengt Hultqvists väg 1, 981 92 Kiruna, Sweden

[a)] Author to whom correspondence should be addressed: f.mazza@tudelft.nl

**ABSTRACT**

We present the first experimental application of coherent Raman spectroscopy (CRS) on the ro-vibrational $v_2$ mode spectrum of methane ($CH_4$). Ultrabroadband femtosecond/picosecond (fs/ps) CRS is performed in the molecular fingerprint region from 1100 to 2000 $cm^{-1}$, employing fs laser-induced filamentation as the supercontinuum generation mechanism to provide the ultrabroadband excitation pulses. We introduce a time-domain model of the $CH_4$ $v_2$ CRS spectrum, including all the five ro-vibrational branches allowed by the selection rules $\Delta v=1$, $\Delta J=0, \pm 1, \pm 2$; the model includes collisional linewidths, computed according to a modified exponential gap scaling law and validated experimentally. The use of ultrabroadband CRS for *in-situ* monitoring of the $CH_4$ chemistry is demonstrated in a laboratory $CH_4$/air diffusion flame: CRS measurements in the fingerprint region, performed across the laminar flame front, allow the simultaneous detection of molecular oxygen ($O_2$), carbon dioxide ($CO_2$), and molecular hydrogen ($H_2$), along with $CH_4$. Fundamental physicochemical processes, such as $H_2$ production via $CH_4$ pyrolysis, are observed through the Raman spectra of these chemical species. In addition, we demonstrate ro-vibrational $CH_4$ $v_2$ CRS thermometry, and we validate it against $CO_2$ CRS measurements. The present technique offers an interesting diagnostics approach to *in-situ* measurement of $CH_4$-rich environments e.g. in plasma reactors for $CH_4$ pyrolysis and $H_2$ production.

## I. INTRODUCTION

Methane ($CH_4$) is the simplest hydrocarbon molecule and one of the most abundant chemical species in the universe, with its ubiquitous chemistry having important implications in planetary[1,2] and life[3] science. In recent years, the challenges posed by anthropogenic climate change have highlighted the critical role played by $CH_4$ in the oil and gas industry and in combustion technologies. $CH_4$ is one of the most powerful greenhouse gases and, despite being short-lived in the Earth's atmosphere, it has the second largest effective climate forcing potential after carbon dioxide ($CO_2$)[4]: $CH_4$ is the major component of natural gas and it's associated with all hydrocarbon fuels. On the other hand, (compressed or liquefied) natural gas is the main alternative fuel for internal combustion engines[5], reducing the emission of transport vehicles with respect to typical diesel and gasoil engines, and $CH_4$ can be produced in a sustainable fashion as a bio-[6] and e-fuel[7]. In this respect, $CH_4$ can also be used as a chemical energy storage to enable the development of the hydrogen economy[7]: $CH_4$ can be transported within the existing infrastructure for natural gas and valorised via chemical reforming to molecular hydrogen[8] ($H_2$) or other commodity hydrocarbons[9,10].

In view of this, $CH_4$ has been extensively investigated by laser diagnostics with linear optical techniques, based on absorption or scattering, providing *in-situ* detection and ro-vibrational spectroscopy in energy systems. Optical probes based on laser absorption have been developed using e.g.

near-[11] and mid-IR[12] diode lasers, quantum-cascade lasers[13], and frequency combs[14], and have been vastly employed for $CH_4$ detection in atmospheric environments[15] and for combustion diagnostics[16]. Absorption techniques have the advantage of simplicity and robustness, as well as the availability of large spectroscopic databases for the interpretation of experimental spectra[17], they can realise ultrafast spectroscopy employing femtosecond (fs) laser sources[18], and the capability for imaging measurements has also been demonstrated[19,20]. On the other hand they suffer from the lack of spatial resolution in the longitudinal direction as they are line-of-site techniques[16], which can severely limit the accuracy of quantitative spectroscopy in inhomogeneous gas-phase media, such as turbulent combustion environments.

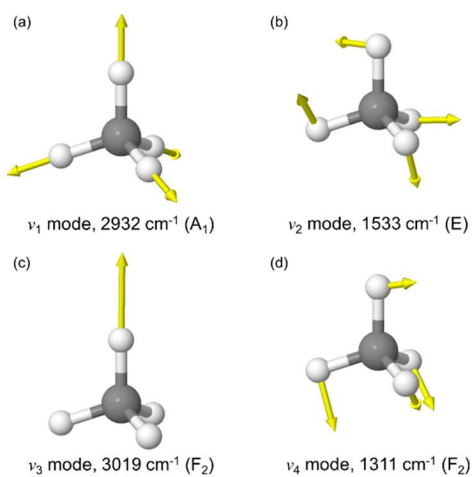

**FIG. 1.** The normal vibrational modes of methane[21]. (a) Symmetric C-H stretch mode. (b) Doubly degenerate H-C-H bend mode. (c) Triply degenerate asymmetric C-H stretch mode. (d) Triply degenerate H-C-H bend mode. (*Jmol: an open-source Java viewer for chemical structures in 3D. http://www.jmol.org/*)

The use of spontaneous Raman scattering in these contexts allows for spatially-resolved measurements, with the possibility for imaging and high repetition rates[22]. Spontaneous Raman spectroscopy has been thoroughly applied to the study of $CH_4$[23,24], both in combustion[25–29] and plasma[21,30] environments, demonstrating its suitability as a probe molecule for the direct measurement of the rotational and vibrational temperature even in non-equilibrium systems, with important implications for $CH_4$ chemistry[31]. The main drawback of spontaneous Raman is the typically low cross section for this incoherent scattering process, which complicates its application in luminous environments, requiring large laser pulse energies to achieve single-shot measurements, with the risk of inducing the optical breakdown of the gas-phase medium or molecular photofragmentation[27]. Additionally, further incoherent emission processes (e.g. chemiluminescence, fluorescence, *etc.*) can shadow the Raman signal, and the use of nanosecond (ns) laser pulses limits the temporal resolution so that spontaneous Raman spectroscopy cannot resolve ultrafast dynamics such as energy redistribution and relaxation processes[32].

The limitations due to the incoherent nature of the Raman scattering process can be overcome by resorting to non-linear optics, in particular in the form of stimulated (SRS) and coherent Raman scattering (CRS). Despite the different nature of their emission[33], SRS and CRS share the use of driving pump and Stokes laser fields with frequency difference matching the ro-vibrational transitions of the Raman-active molecules, resulting in the emission of a coherent signal, which can be remotely detected even in harsh environments. Since the very beginning of their development, continuous-wave SRS[34,35] and CRS[36–39] have been employed to perform high-resolution measurements of the ro-vibrational Raman spectrum of $CH_4$ in the so-called pentad region, allowing for the assignment of individual rotational lines in the isotropic Q-branch of the symmetric C-H stretch ($v_1$ mode, see **FIG. 1(a)**). They were further employed to investigate the rotational energy transfer in molecular collisions[40–42] between the vibrationally coherent $CH_4$ molecules and other collisional partners, e.g. $N_2$ and argon (Ar). In the context of combustion diagnostics, the $CH_4$ CRS signal was recorded at temperatures as high as 1273 K in a furnace[43], and $CH_4$ ns CRS was used to measure the temperature in a laminar $CH_4$/air diffusion flame[44] and in supercritical LOX/$CH_4$ combustion[45].

In recent years, the introduction of ultrafast laser sources, providing pulses of the duration of pico- (ps) and femtosecond (fs), has led to the development of various kind of time-resolved CRS techniques, able to directly measure the rotational and vibrational wave packets of Raman-active molecules[46]. In particular, hybrid fs/ps CRS[47], employing a combination of broadband fs pump and Stokes and narrowband ps probe pulses, achieved simultaneous time- and frequency resolution, measuring the CRS spectrum below the relaxation timescale of the ro-vibrational energy. Different kinds of time-resolved CRS have been applied to $CH_4$. Relative $CH_4$ concentrations measurements in binary mixtures with $N_2$ were demonstrated via hybrid CRS by Engel *et al.*[48] and via chirped-probe-pulse (CPP) CRS by Dennis *et al.*[49]. Bohlin and Kliewer at Sandia National Laboratories (SNL), Livermore, CA, employed a hollow-core fibre to compress a 45 fs pulse to <7 fs supercontinuum to excite the ro-vibrational Raman modes up to the pentad region of the $CH_4$ Raman spectrum, realising ultrabroadband fs/ps CRS imaging[50]. In a later work[51], the same group employed ultrabroadband CRS imaging to investigate $CH_4$/air flame-wall interaction, realising the simultaneous detection of $N_2$, $H_2$, molecular oxygen ($O_2$), carbon monoxide (CO), $CO_2$, and $CH_4$. Recently, researchers at SNL Livermore and at the Plasma Physics Laboratory, Princeton, NJ, have presented a thorough investigation of the $v_1$ mode spectrum of by hybrid fs/ps CRS[52]: they demonstrated the viability of $CH_4$ CRS thermometry up to 1000 K, realised time-resolved measurements of the collisional dephasing of the Raman coherence of $CH_4$ perturbed by itself, by $N_2$ and by Ar, and developed a time-domain model of the $CH_4$ $v_1$ CRS spectrum.

All the applications of CRS to $CH_4$ have been focused on the pentad region of its spectrum (~2600-3300 cm$^{-1}$), which is dominated by the fundamental bands of the $v_1$ mode and of the asymmetric C-H stretch ($v_3$) mode (see **FIG. 1(c)**). The so-called dyad region (~1200-1900 cm$^{-1}$), due to the doubly and triply degenerate bending of the H-C-H bonds[23] ($v_2$ and $v_4$ modes, in **FIG. 1(b)** and **FIG. 1(d)** respectively), has been thus far neglected as the Raman cross section of the bending modes is more than one order of magnitude smaller than that of the stretch modes. On the other hand, this represents an important region for combustion and plasma diagnostics, as it overlaps with the ro-vibrational spectra of $CO_2$[53] and $O_2$[54], as well as with the pure-rotational spectrum of $H_2$, thus offering an interesting window for the study of $CH_4$ chemistry with high spatio-temporal resolution. In view of this we here present the first CRS investigation of the $v_2$ mode of $CH_4$, dominating the dyad region of its Raman spectrum. We employ a two-beam hybrid fs/ps CRS system[55] to perform ultrabroadband CRS spectroscopy in the region 1100-1200 cm$^{-1}$ of the Raman spectrum, using *in-situ* fs-laser-induced filamentation as the compressed supercontinuum generation mechanism[54,56,57]. The use of hybrid fs/ps allows us to simultaneously access the frequency and time-domain to investigate the coupling of the rotational and vibrational energy degrees of freedom of the $CH_4$ molecule, as well as the rotational energy transfer in molecular collisions. In order to demonstrate the potential of $CH_4$ $v_2$ CRS as a diagnostic tool for combustion and plasma technologies, we performed spatially-resolved multiplexed measurements across a laminar $CH_4$/air diffusion flame.

## II. METHODS

### A. Optical setup

The experimental setup for two-beam ultrabroadband fs/ps CRS is similar to the one detailed in a previous work[54], only a brief summary is here provided. A regenerative chirped-pulse amplifier (CPA) system (Astrella, Coherent) is employed as the single light source, providing fs laser pulses at 800 nm, with a pulse energy of ~7.5 mJ, at a repetition rate of 1 kHz. A 65% split of the total pulse energy is temporally compressed to 35 fs and fed to a second-harmonic bandwidth compressor (SHBC, Light Conversion), to generate a frequency-doubled ~5 ps pulse, which serves as the narrow-band probe pulse. Its duration and spectral linewidth are tuned by a 4f-pulse shaper to 7.3 ps and 4.1 cm$^{-1}$ (time-bandwidth product: 0.89 for a top-hat pulse[58]), with a resulting probe pulse energy of 200 µJ. The remainder of the CPA output is employed as a single degenerate fs pump/Stokes pulse in a two-beam phase CRS matching configuration[55,59]: an external compressor unit provides independent control the spectral phase of the fs pulse, so as to compensate for the dispersion due to the optical elements on the pump/Stokes beam path. As demonstrated in previous works[54,60], fs laser-induced filamentation can be used to generate a compressed supercontinuum, which acts as a single ultrabroadband pump/Stokes pulse, coherently exciting all the Raman-active modes whose characteristic period is longer than the supercontinuum pulse. Here this *in-situ* generation/*in-situ* use scheme is employed to perform ultrabroadband CRS in the vibrational fingerprint region: a spherical lens (f:=500 mm) focuses the ~1.2 mJ pump/Stokes pulse, with a theoretical irradiance of ~357 TW/cm$^2$ at the focal point, thus resulting in the formation of a laser-induced plasma filament, which propagates for ~13 mm before collapsing. The laser focusing lens is mounted on a linear stage, whose position is adjusted to maintain the filament ~4 mm before the probe volume. As the fs pulse experiences self-phase modulation and non-linear propagation in the plasma medium inside the filament, these result in a compressed supercontinuum output, which serves as the ultrabroadband pump/Stokes pulse.

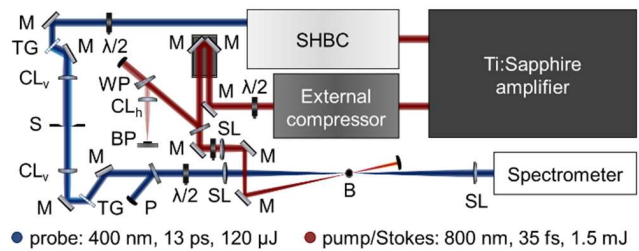

**FIG. 2.** Schematic of the ultrabroadband two-beam fs/ps CRS instrument. A single Ti:sapphire regenerative amplifier is employed to generate the 35 fs pump/Stokes and the 7 ps probe pulses. *In-situ* fs filamentation is employed as the supercontinuum generation mechanism: the ultrabroadband pump/Stokes beam is crossed by the probe beam at a distance of ~4 mm after exiting the filament. TG, transmission grating; S, slit; TS, translation stage; CL, cylindrical lens with horizontal (h) and vertical (v) symmetry axes; λ/2, half-wave plate; WP, wedge plate; BP, beam profiler; B, burner; P, polariser; SL, spherical lens; M, mirror.

A crossed beam quasi-phase-matched configuration is adopted with the probe beam focused by a spherical lens (f:=300 mm) to the measurement location and crossing the pump/Stokes at a ~3° angle, resulting in an estimated 20 μm (width, FWHM) x 2.5 mm (length, FWHM) x 20 μm (height, FWHM) probe volume. The relative delay of the probe pulse is tuned by controlling an automated delay line on the pump/Stokes beam path with a <10 fs resolution (Thorlabs). A combination of 800 nm half-wave plate and thin-film polariser is employed to split a ~50 μJ portion of the beam: a single reflection from a wedge plate is focused by a cylindrical lens (f:=300 mm) onto a sCMOS beam profiler (WinCamD, Dataray), so as to monitor the (most sensitive) vertical alignment of the pump/Stokes beam over the whole range of motion of the automated stage. The ultrabroadband CRS signal is collimated through a spherical lens (f:=400 mm) and a band-pass filter (20 nm bandwidth FWHM, Semrock) is employed to suppress the probe, almost co-propagating with the signal in this geometry. The signal is finally dispersed through a high-resolution transmission grating (3039 lines/mm, Ibsen Photonics) and imaged by a relatively fast-focusing lens (f:=200 mm) onto the sCMOS detector (Zyla, Andor), resulting in a detection bandwidth of 900 cm$^{-1}$, with a dispersion of 0.43 cm$^{-1}$/pixel. The 4.1 cm$^{-1}$ spectral linewidth of the ps probe pulse is the limiting factor for the resolution of the ultrabroadband CRS signal; the instrumental broadening due to the spectrometer is evaluated by fitting a Voigt profile with ~0.1 cm$^{-1}$ FWHM Lorentzian and ~1.7 cm$^{-1}$ FWHM Gaussian contributions. Half-wave plates are employed to maximise the efficiency of the transmission grating (measured to >90% for S-polarisation at 400 nm) in the 4f-filter and to control the relative polarisation angle between the pump/Stokes and the probe field: in the present work the beams were set to have same polarisation, so as to maximise the signal generation efficiency, and a 400 nm half-wave plate is employed to turn the polarisation of the (completely depolarised) CH$_4$ $\nu_2$ CRS signal and optimise its diffraction efficiency in the spectrometer.

**B. Time-resolved CRS measurements**

The $\nu_2$ mode Raman spectrum of CH$_4$ is here investigated under a number of different experimental conditions, by varying the collisional environment to measure its dephasing coefficients, and by performing ultrabroadband spectroscopy in a CH$_4$/air flame, demonstrating single-shot detection of the CH$_4$ $\nu_2$ CRS spectrum at temperatures as high as ~800 K.

The collisional dephasing of the CH$_4$ $\nu_2$ CARs signal was measured by performing time-resolved measurements of its spectrum and delaying the probe pulse relative to the pump/Stokes up to 220 ps, in steps of 5.1 ps. Samples of 1000 single-shot CH$_4$ $\nu_2$ CRS spectra were recorded up to ~80-100 ps, where the collisional dephasing results in a reduction of the signal-to-background ratio (SBR) to ~3.3, and 10-shot-averaged spectra were recorded from that point onward: at the transition point both single-shot and shot-averaged datasets were acquired and compared, computing a scaling factor over the different dynamic ranges of the dephasing experiments. Collisional dephasing measurements were performed, at atmospheric pressure, in different mixtures of combustion-relevant gases. The measurements were performed in an open flow, regulated by digital flow controllers (Bronkhorst), provided inside a stainless steel T-junction: two set of measurements for binary mixtures of 75% and 50% CH$_4$ in combination with N$_2$, H$_2$ and argon were performed. The demonstration of the single-shot detection of the CH$_4$ $\nu_2$ CRS spectrum at high temperatures in a chemically-reacting flow was realised in a laminar CH$_4$/air flame, provided on a Bunsen burner. The burner is a seamless steel pipe with an inner diameter of 19 mm, through which the CH$_4$ flow was delivered with an exit plane velocity of 2.65 cm/s, ensuring a laminar flow (Reynolds number ~35); a steel mesh was placed ~15 mm above the burner to stabilise the flame, while the measurements were performed ~1 mm above its nozzle. Point-wise ultrabroadband CRS measurements were performed at 25 locations across the flame front, moving from the centre of the burner (location y=0 mm) with a 0.5 mm step size. In order to extend the detection limit at lower concentrations and higher temperatures, a set of 1000 10-shot-averaged CRS spectra was acquired at each measurement location.

**C. Time-domain CRS model**

The modelling of the CH$_4$ $\nu_2$ coherent Raman spectrum follows the modelling approach of the CH$_4$ $\nu_1$ spectrum presented by T. Chen et al.[52], based on assumption of impulsive excitation of the Raman coherence, and on a phenomenological description of the resulting third-order nonlinear optical susceptibility following Prince et al.[47] This approach has been successfully demonstrated for both pure-rotational[61–63] and ro-vibrational[54,64] CRS of multiple combustion-relevant species. The main assumptions of the time-domain model can be outlined as follows: the nonlinear interaction of the electronically off-resonant pump, Stokes probe laser fields (at angular frequencies $\omega_i$ with i=1, 2, and 3, for the pump, Stokes and probe) with the gas-phase medium induces the macroscopic third-order polarisation:

$$P^{(3)}(t,\tau_{12},\tau_{23}) = \left(-\frac{i}{\hbar}\right)^3 \int_{-\infty}^{t} dt_3 \int_{-\infty}^{t_3} dt_2 \int_{-\infty}^{t_2} dt_1 \Big[ E_3(t-t_3) e^{i(\omega_3-\omega_2+\omega_1)} \cdot$$
$$\delta(t_3) \chi^{(3)}(t_2) E_2^*(t-t_3+\tau_{23}-t_2) e^{i(-\omega_2+\omega_1)} \cdot \qquad (1)$$
$$E_1(t-t_3+\tau_{23}-t_2+\tau_{12}-t_1) e^{i\omega_1} \delta(t_1)$$

where $E_i$ and $E_i^*$ are the envelope of the i$^{th}$ field and its complex conjugate, $t_i$ is the coherence timescale for the molecules after the interaction with the i$^{th}$ field, $\tau_{ij}$ is the delay of the interaction with the i$^{th}$ and j$^{th}$ fields, $\delta$ is the Dirac function, representing the instantaneous dephasing of the (virtual) electronic coherence, and $\chi^{(3)}$ is the third-order susceptibility of the gas-phase medium. Under the assumption of impulsive excitation of the ro-vibrational Raman coherence we can disregard the temporal envelope of the combined pump/Stokes field (in case of two-beam CRS[55]) and write:

$$P^{(3)}(t,\tau_{23}) = \left(-\frac{i}{\hbar}\right)^3 E_3(t+\tau_{23}-t_2) \chi^{(3)}(t_2) \qquad (2)$$

Hence the CRS intensity spectrum is obtained, upon Fourier transforming the polarisation field, as:

$$I_{CRS}(\omega,\tau_{23}) = \left| \mathcal{F}\{P^{(3)}(t,\tau_{23})\} \right|^2 \qquad (3)$$

$\chi^{(3)}$ represents the core of the model outlined so far, and is typically treated phenomenologically as:

$$\chi^{(3)}(t_2) = \sum_{(v,J)} W_{(v_i,J_i)\to(v_f,J_f)} \exp\left[\left(i\omega_{(v_i,J_i)\to(v_f,J_f)} - \Gamma_{(v_i,J_i)\to(v_f,J_f)}\right)t\right] \qquad (4)$$

with summation running over all the possible initial and final ro-vibrational states (represented by the total angular momentum and vibrational quantum numbers, $v$ and $J$) allowed by the Raman selection rules for the molecular species considered. The temporal evolution of $\chi^{(3)}$ is thus computed as the interferogram of damped harmonic oscillations –at the Raman frequencies $\omega_{(v_i,J_i)\to(v_f,J_f)}$ and with damping coefficients $\Gamma_{(v_i,J_i)\to(v_f,J_f)}$ dominated by the collisional rotational energy transfer (RET) at atmospheric pressure[65]–, weighted by the semi-classical transition probabilities[66]. Eq. (4) thus introduces in the time-domain CRS model the spectroscopic properties of the Raman-active species, which in turn depend on the chemical structure of the molecule considered.

The case of CH$_4$ is considerably more complicated than simple diatomic molecules, as early studies of its absorption spectrum revealed[67]. CH$_4$ is a polyatomic spherical top molecule, possessing three 3-fold and 2-fold axes of symmetry, and six planes of symmetry: in terms of these symmetry elements, CH$_4$ belongs to the tetrahedral point group (T$_d$)[68]. Since the molecule has five nuclei, there are nine possible vibrational modes which, according to group theory, can be clustered in the four normal modes illustrated **FIG. 1**, called by the Greek letter "$v$" and identified by the vibrational quantum numbers (v$_1$, v$_2$, v$_3$, v$_4$) and by the symmetry labels A, E, and F, for non-degenerate, doubly degenerate and triply degenerate modes, respectively. The first normal mode ($v_1$ at 2932.4 cm$^{-1}$), as already mentioned, is the non-degenerate symmetric stretch of the C-H bonds, which preserves the symmetry of the molecule in the vibrational ground state (0000), so that the instantaneous polarisability tensor has zero anisotropic invariant, and the associated Raman spectrum consists only of the isotropic Q-branch (with selection rules: $\Delta v=1$, $\Delta J=0$). The $v_2$ mode (at 1533.3 cm$^{-1}$) is a doubly degenerate bend of the H-C-H bonds: this mode, as all the other vibrational modes of CH$_4$ is Raman active, with selection rules $\Delta J=0, \pm 1, \pm 2$, so that the Raman spectrum is composed of O-, P-, Q-, R- and S-branch lines. The asymmetric C-H stretch mode ($v_3$ at 3018.5 cm$^{-1}$) and the $v_4$ bend mode (at 1310.8 cm$^{-1}$), being the least symmetric modes, are triply degenerate and the reduced symmetry results in the selection rules: $\Delta J=0, \pm 1, \pm 2$. Experimental measurements on the depolarisation ratio of the fundamental $v_2$ and $v_3$ Raman spectra have shown that these modes are completely depolarised (i.e. $\rho=0.75$), with the Raman activity entirely due to the anisotropic part of the polarisability tensor. The normal frequencies of the stretch modes are roughly double than the bending modes (i.e. $v_1 \approx v_3 \approx 2v_2 \approx 2v_4$), so that vibrational bands can be grouped into polyads, denoted $P_n$, with polyad number $n = (2v_1 + 2v_3 + v_2 + v_4)$. $P_0$ is the vibrational ground state; $P_1$ has two vibrational levels, and Raman transitions $P_0 \leftarrow P_1$ involving the fundamental bands of $v_2$ and $v_4$ constitute the dyad region of the Raman spectrum; similarly, $P_2$ has five vibrational levels, and transitions $P_1 \leftarrow P_2$ involve the fundamental $v_1$ and $v_3$ modes, as well as the overtones (2$v_2$ and 2$v_4$) and combination ($v_2+v_4$) of the bending modes, defining the pentad region.

The concept of symmetry plays an essential role in molecular spectroscopy, as it constraints the form of the molecular wave function according to Pauli's exclusion principle. Under the assumption of the Born-Oppenheimer approximation, the total wave function of the CH$_4$ molecule is written as the product of separable functions:

$$\psi = \psi_T \psi_E \psi_V \psi_R \psi_S \qquad (5)$$

with $\psi_T$ translational wave function, $\psi_E$ electronic wave function (including the electron spin), $\psi_V$ vibrational wave function, $\psi_R$ rotational wave function, and $\psi_S$ nuclear spin wave function. As the translational motion doesn't involve any internal degree of freedom, $\psi_T$ is completely symmetrical (A symmetry), and so is $\psi_E$ in the electronic ground state. The symmetry of the vibrational wave function depends on the symmetry of the irreducible representations of the $T_d$ point group corresponding to the normal vibrational modes in **FIG. 1**. The rotational state of a $CH_4$ molecule can be specified by the quantum numbers $J, M$, where $M$ represents the projection of the total angular momentum with respect to an inertial frame of reference: the associated wave function is a solution of the spherical top wave equation. E.W. Wilson Jr. showed that this solution is a linear combination of irreducible representations of symmetry A, E and F[69]: his results are summarised in **TABLE 1**. The rotational wave function has $(2J+1)$ degenerate projections with respect to an arbitrary inertial axis (i.e. $M = 0, \pm 1, \ldots, \pm J$), hence the common factor in the table: this degeneracy cannot be lifted by intramolecular interaction, but only by application of an external field (e.g. in Stark effect)[68].

**TABLE 1** Symmetry of the rotational wave function of $CH_4$ (with $p = 0$, 1, 2, 3, 4, or 5).

| | |
|---|---|
| $J = 6p$   | $(2J+1) [ (p+1)A + 2p\,E + 3p\,F ]$ |
| $J = 6p+1$ | $(2J+1) [ p\,A + 2p\,E + 3(3p+1)\,F ]$ |
| $J = 6p+2$ | $(2J+1) [ p\,A + 2(p+1)\,E + 3(3p+1)\,F ]$ |
| $J = 6p+3$ | $(2J+1) [ (p+1)A + 2p\,E + 3(3p+2)\,F ]$ |
| $J = 6p+4$ | $(2J+1) [ (p+1)A + 2(3p+1)\,E + 3(3p+2)\,F ]$ |
| $J = 6p+5$ | $(2J+1) [ p\,A + 2(p+1)\,E + 3(3p+3)\,F ]$ |

Finally, $CH_4$ has 16 possible nuclear spin wave functions: 2 are singlet states (with nuclear spin quantum number $I = 0$, and associated projection $M_I = 0$), 3 are triplet states ($I = 1$, $M_I = 0, \pm 1$) and one is a quintet states ($I = 2, M_I = 0, \pm 1, \pm 2$). In terms of symmetry, $\psi_S$ is the combination[69]: $5A + E + 3F$. The $CH_4$ molecule has 4 identical $^1H$ nuclei, which obey the Fermi-Dirac statistics: according to Pauli's principle, the total wave function must be symmetric with respect to any proper rotation of the $T_d$ point group, so that its overall symmetry must be $A$.

**TABLE 2** Multiplication table for the irreducible representations.

|   | A | E | F |
|---|---|---|---|
| A | $A = A$ | $E \not\supseteq A$ | $F \not\supseteq A$ |
| E | $E \not\supseteq A$ | $(2A + E) \supseteq A$ | $F \not\supseteq A$ |
| F | $F \not\supseteq A$ | $F \not\supseteq A$ | $(A + E + 2F) \supseteq A$ |

Under the multiplication in Eq. (5) the symmetry label obeys the rules in **TABLE 2**, and only wave functions with resulting symmetry $A$ are allowed. This has important implications for the statistical weights of the ro-vibrational energy levels included in the calculation of Boltzmann distribution:

$$\rho = g_V g_R g_S \frac{\exp(-\hbar c E_{v,J}/k_B T)}{Z} \quad (6)$$

where $E_{v,J}$ is the ro-vibrational energy (expressed in cm$^{-1}$) of the molecule ($v$ is a short-hand notation for the complete set of vibrational quantum numbers), $g_V$ is the vibrational degeneracy, $g_R$ is the degeneracy of the rotational wave function according to the entries of **TABLE 1**, and $g_S$ is the nuclear spin degeneracy, being 5, 2 and 3 for $A$, $E$ and $F$ symmetry labels, respectively. According to the symmetry multiplication rules in **TABLE 2** only the product of identical symmetry labels produces a representation with $A$ symmetry: hence the often statement, in the literature, that ro-vibrational states (i.e. $\psi_V \psi_R$) of symmetry $A$, $E$, and $F$ are 5-, 2- and 3-fold degenerate.

In order to model the temporal evolution of the nonlinear susceptibility according to Eq. (4), one must know the frequencies of the allowed Raman transitions, and the corresponding polarisabilities of the $CH_4$ molecule, which enter the weighting factors as[44]:

$$W_{(v_i,J_i) \to (v_f,J_f)} = |\langle v_i, J_i | \hat{\alpha} | v_f, J_f \rangle|^2 \left( \rho_{(v_f,J_f)} - \rho_{(v_i,J_i)} \right) \quad (7)$$

These quantities can be computed adopting a tetrahedral formalism developed for the analysis of spherical top molecules[70], which removes inter-polyad contributions to the higher order thanks to implicit contact transformations[71], while explicitly retaining inter-polyad contributions, due to the strong coupling between vibrational states within individual polyads. This approach was developed at the University of Burgundy and allowed the derivation of effective ro-vibrational Hamiltonians for polyads up to the tetradecad (i.e. $P_4$, at ~6200 cm$^{-1}$), from which the position and intensities of the spontaneous Raman spectrum of $CH_4$ are computed[72–74]. The resulting spectral database has been employed by multiple research groups to simulate the spontaneous $CH_4$ Raman spectrum in both the dyad[75] and pentad region[21,30,76], and recently its application has been extended to the $v_1$ $CH_4$ CRS spectrum[52]. According to the notation of Jourdanneau et al. the spontaneous Raman line intensities in the spectral database are computed as[76]:

$$I_{(v_i,J_i) \to (v_f,J_f)} \propto \rho_{(v_i,J_i)} \left[ S_0 \mathcal{J}^2_{(v_i,J_i) \to (v_f,J_f)} + S_2 \mathcal{A}^2_{(v_i,J_i) \to (v_f,J_f)} \right] \quad (8)$$

where $S_{0,2}$ are the Stone coefficients, depending on the observation geometry in a spontaneous Raman experiment, while $\mathcal{J}$ and $\mathcal{A}$ are the transition values of the isotropic and anisotropic components of the polarisability tensor. As already mentioned, the $v_2$ mode Raman spectrum is completely depolarised, so that its transition polarisability has only an anisotropic component, and the corresponding Stone coefficient become an irrelevant scaling factor. In contrast to the incoherent spontaneous Raman scattering, the probability amplitude for the coherent Raman scattering process depends on the differential of the Boltzmann population in the initial and final ro-vibrational states, as shown in Eq. (7). The weight factors in Eq. (4) are thus computed from the line intensities in the spectral database as:

$$W_{(v_i, J_i) \to (v_f, J_f)} = I_{(v_i, J_i) \to (v_f, J_f)} \left( \frac{\rho_{(v_f, J_f)} - \rho_{(v_i, J_i)}}{\rho_{(v_i, J_i)}\big|_{T_0}} \right) \quad (9)$$

with $T_0$ (~1500 K) being the reference temperature for the calculation of the line intensities in the spectral database[21]. The spectral database contains ~16 million lines in the dyad region – involving vibrational states up to the tetradecad and rotational states up to $J = 23$: of these ~11 million correspond to $v_2$ mode Raman transitions (i.e. $\Delta v_2 = 1$, $\Delta v_4 = 0$) and ~5 million are $v_4$ mode transitions. The $v_4$ mode band is the weakest of the ro-vibrational Raman spectrum of CH$_4$ and could not be observed in our CRS experiments: the time-domain CH$_4$ CRS model was thus limited to the $v_2$ mode spectrum. Each of the $v_2$ mode transitions gives rise to a damped harmonic contribution which is computationally evaluated on a temporal grid up to 1 ns, with 8 fs step size, corresponding to a resolution of 0.03 cm$^{-1}$ for the Fourier-transformed CRS signal. As pointed out by Chen *et al.*[52], the main challenge in the development of an accurate time-domain model for the CH$_4$ CRS spectrum is the staggering number of Raman transitions included in the MeCaSDa calculated spectral database computed at the University of Burgundy[72,73] (and accessible at the following url: http://vamdc.icb.cnrs.fr/PHP/methane.php) and the resulting computational cost. Implementing the calculation of $\chi^{(3)}$ as a running sum for every transition input avoids exceeding memory limits of the available computational resources; each spectral calculation still needs to be iterated for different input values of temperature and, if the effects of molecular collisions are not negligible, of species concentrations in the probe volume. Hence, the calculation of spectral libraries for quantitative spectroscopy can be a cumbersome task, especially when the possibility of CH$_4$ CRS thermometry in non-equilibrium environments is considered[21]. The computational cost can be reduced by implementing a filter to exclude transitions whose intensity is lesser than the a cut-off value of strongest transition in the spectrum[52], depending on both the strength of the associated instantaneous dipole and on the Boltzmann population.

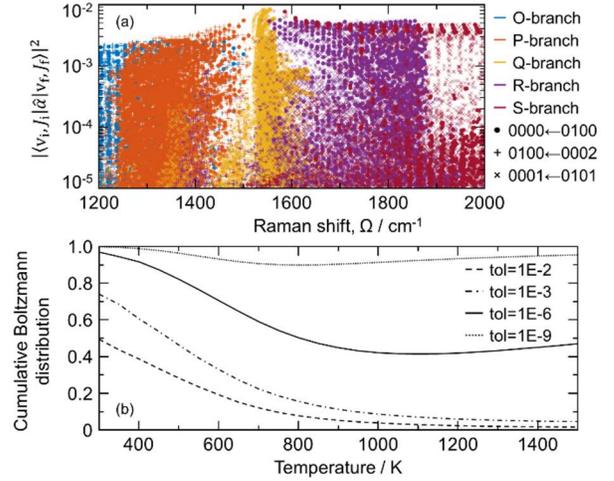

**FIG. 3.** The spontaneous Raman cross-section for CH$_4$ $v_2$ mode transitions. (a) Transition moment of the polarisability anisotropy for the fundamental band (0000←0100) and the first two hot bands (0100←0200 and 0001←0101) of the $v_2$ mode, according to the different Raman selection rules. (b) The cumulative distribution function, representing the fraction of the Boltzmann population included in the calculation of $\chi^{(3)}$ as a function of the input temperature and cut-off value selected for the filtering of the spectral database.

**FIG. 3(a)** shows the transition polarisability for a small fraction of the Raman transitions considered in the database, corresponding to the fundamental band of the $v_2$ mode spectrum (0000←0100) and its first two hot bands (0001←0101 and 0100←0002): the overall database spans more than 18 orders of magnitude in the Raman polarisability. These are combined with the differential Boltzmann population between the initial and final transition states at different temperatures, between 300 and 1500 K in **FIG. 3(b)**, to compute the overall line intensities of the CRS spectrum. Implementing different cut-off values the number of Raman transitions taken into account in the calculation of $\chi^{(3)}$ can be significantly reduced, at cost of considering only a fraction of the Boltzmann population. In particular, adopting a tolerance of 1% or 0.1% of the strongest line in each branch most of the cumulative distribution function (CDF) is neglected. Increasing the tolerance to a millionth (billionth) of the strongest line, more than 40% (90%) of the CDF is taken into account at all temperatures considered in the theoretical CRS library. The non-monotonic behaviour shown by the corresponding curve in **FIG. 3(b)** depends on the combination of the different Raman polarisabilities and of the spreading Boltzmann distribution at higher temperatures. A cut-off value of 1E-4 was found to be sufficient to model the time-domain behaviour of the CH$_4$ $v_2$ CRS spectrum at room temperature, where the effect of vibrational hot bands is measured to be negligible. In the flame experiment, on the other hand, the cut-off value needs to be increased to guarantee the independence of $CH_4$ $v_2$ CRS thermometry from the spectral filtering. An unfiltered spectral library was used to fit the flame spectra up to ~800 K, and a cut-off value of 1E-6 was then found to perfectly reproduce these results while including only half of the cumulative Boltzmann distribution as shown in **FIG. 3(b)**.

The time-domain $CH_4$ $v_2$ CRS model is complemented by the inclusion of Raman linewidths, representing the collisional dephasing of the signal. The collisional dephasing coefficients of the Q-branch spectrum $\Gamma_{(v_i,J)\to(v_f,J)}$ are modelled according to the modified exponential gap (MEG) law[77,78], as:

$$\Gamma_{(v_i,J)\to(v_f,J)} = \sum_{k\neq j}\gamma_{kj} \quad (10)$$

with upward and downward collisional transition rates, between the $i^{th}$ and $j^{th}$ rotational energy states, given by:

$$\gamma_{ji} = \alpha p \left(\frac{T_0}{T}\right)^n \left(\frac{1+aE_i/k_BT\delta}{1+aE_i/k_BT}\right)^2 \exp\left[\frac{-\beta(E_j-E_i)}{k_BT}\right] \quad (11)$$

$$\gamma_{ij} = \frac{2J_i+1}{2J_j+1}\gamma_{ji}\exp\left(\frac{E_j-E_i}{k_BT}\right) \quad (12)$$

where $p$ is the pressure, $T_0$=296 K is the reference temperature, $E_i$ and $E_j$ represent the ro-vibrational energy in the upper and lower vibrational states (so that the energy gap is always positive), and $\alpha$, $\beta$, $\delta$, $a$ and $n$ are fitting parameter for the scaling law. In the present linewidth model the effect of the symmetry of ro-vibrational wave function is neglected[42], so that the linewidth computed for a specific value of the rotational quantum number $J$ is applied to all its symmetry components in **TABLE 1**; similarly, the same RET rate is assumed for all the vibrational hot bands. Following the work of Chen et al.[52] the species-specific constant $a$ is set to 2, and the number of fitting parameter to be simultaneously determined is reduced by independently fitting $\alpha$ and $\beta$ at room temperature, assuming $\delta$=1 and $n$=0. While the MEG model has been successfully applied to isotropic $v_1$ Q-branch CRS spectrum of $CH_4$ in multiple studies[44,45,52], the sum rule in Eq. (10) is not rigorously satisfied in the case of the anisotropic $CH_4$ $v_2$ lines, where molecular reorientation and degenerate levels are present[79]. To a first approximation this fact can be neglected, and the corresponding Raman lines can be assumed to satisfy Eq. (10): this approximation, though coarse, was employed with satisfactory results to model the collisional line broadening of the P- and R-branch lines in the absorption spectra of $CO_2$[80]. We justify the present use of the MEG scaling law as a first-approximation model of the collisional dephasing of the anisotropic $CH_4$ $v_2$ Q-branch spectrum in view of the fact that only atmospheric pressure experiments are here reported, so that the molecular reorientation in inelastic collisions and the intra-branch coupling can be reasonably neglected. On the other hand, the sum rule in Eq. (10) is not directly used to compute the dephasing coefficients of the O-, P-, R- and S-branch lines, as the inter-branch coupling is assumed to dominate over the intra-branch contribution[79]. The collisional dephasing rates of these lines are computed assuming the "random phase approximation" (RPA), whereby the width of an anisotropic line involving a change in the rotational quantum number ($J_i \to J_f$) depends only on the relaxation of the rotational energy levels labelled by the quantum numbers $J_i$ and $J_f$, as[81]:

$$\Gamma_{(v_i,J_i)\to(v_f,J_f)} = \frac{1}{2}\left(\Gamma_{(v_i,J_i)\to(v_f,J_i)} + \Gamma_{(v_i,J_f)\to(v_f,J_f)}\right) \quad (13)$$

## III. RESULTS AND DISCUSSION

### A. Time-resolved $CH_4$ $v_2$ spectrum

An example of a single-shot $CH_4$ $v_2$ coherent Raman Stokes spectrum (CSRS), acquired in a room-temperature methane flow, in the spectral range 1100-2000 cm$^{-1}$ is given in **FIG. 4**. The $CH_4$ $v_2$ spectrum presents all five branches corresponding to the selection rules $\Delta J$=0, ±1, and ±2. The spectral resolution of our CRS instrument (limited by the 4.1 cm$^{-1}$ width of the probe spectrum) is insufficient to resolve the rotational structure of the Q-branch, which is then a single convolved feature at ~1535 cm$^{-1}$. The ordering of the branch labels with the Raman shift on the Stokes of the probe line is opposite to the one on the anti-Stokes side (CARS), as shown in **FIG. 4**: the O- and P-branch appearing at larger Raman shifts than the R- and S-branch.

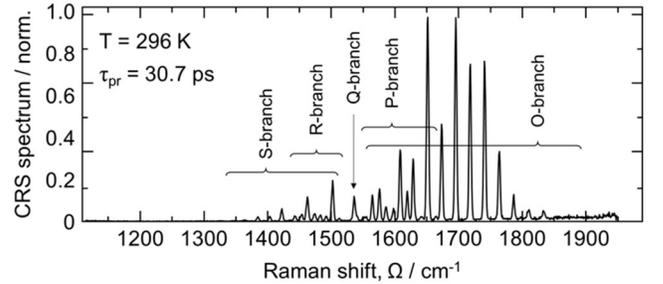

**FIG. 4.** Single-shot $CH_4$ $v_2$ CSRS spectrum at room-temperature. The $v_2$ mode Raman-activity is completely anisotropic and the selection rules allow for O-, P-, Q-, R-, and S-branch transitions. Note that, as the CRS signal is acquired in the Stokes side, the usual branch ordering is reverted, with negative changes in the total angular momentum quantum number ($\Delta J$=-1,-2 for the P- and O-branch, respectively) determining a larger frequency transition.

The spectroscopic data for the $CH_4$ Raman spectrum discussed in Section II can be directly applied to the CSRS

spectrum, with the only caveat that the branch labels for negative and positive changes in $J$ need to be interchanged. A comparison of the experimental measurements and model prediction of the dynamic behaviour of the $CH_4$ $v_2$ CRS spectrum is provided in **FIG. 5**.

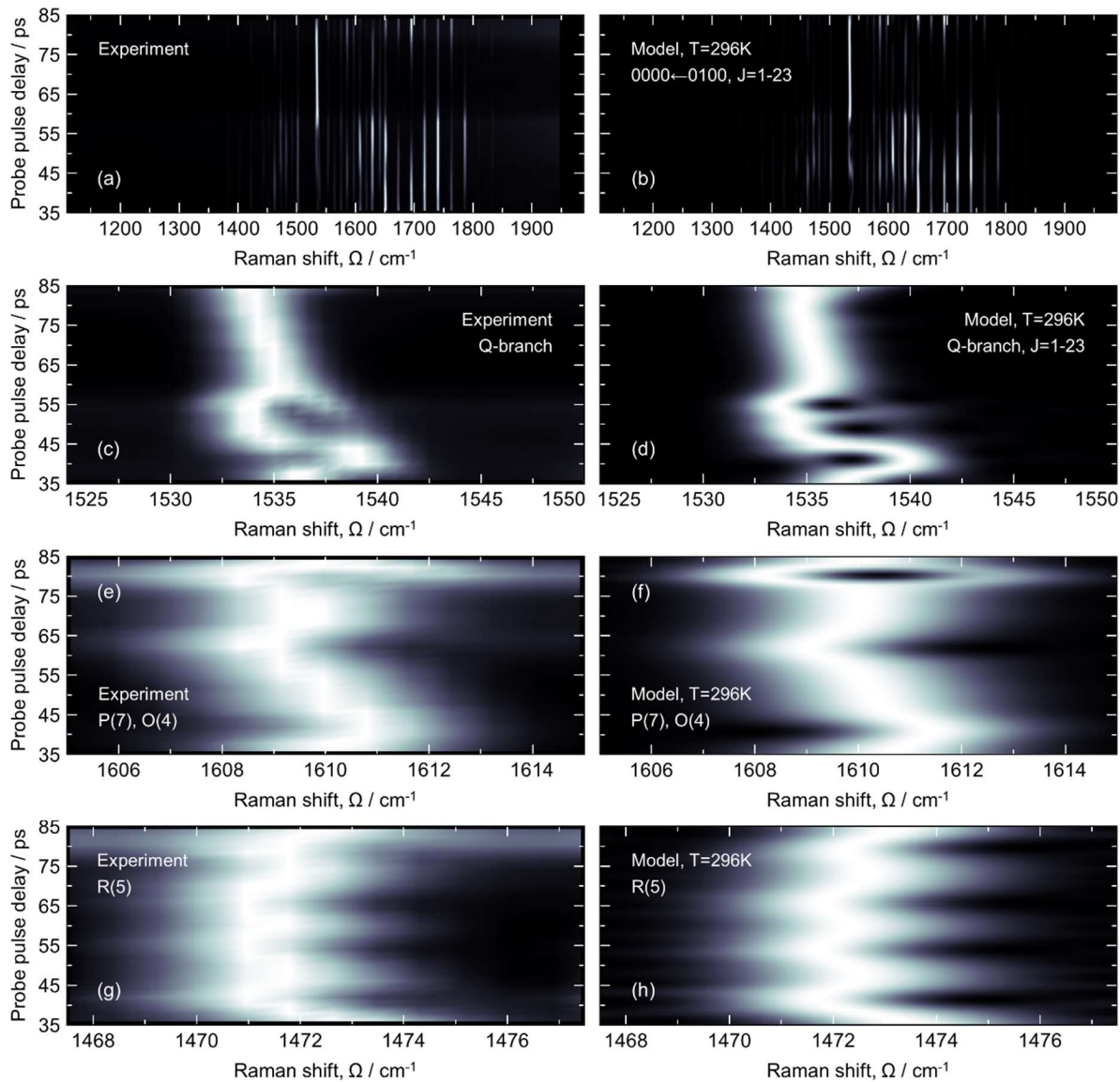

**FIG. 5.** Time-resolved $CH_4$ $v_2$ CRS spectra: comparison between experiments and time-domain model. (a) Experimental spectrochronogram of the ro-vibrational $CH_4$ $v_2$ mode, acquired in a room-temperature $CH_4$ flow over for values of the probe pulse delay spanning over 50 ps. A 1000-shot-averaged CRS spectrum is shown at each probe delay. (b) Modelled spectrochronogram including only the fundamental band of the $v_2$ mode (i.e. 0000←0100), and rotational states up to J=23. (c) Experimental spectrochronogram of the $v_2$ Q-branch spectrum. (d) Modelled spectrochronogram of the $v_2$ Q-branch spectrum. (e) Experimental spectrochronogram of the convoled P(7) and O(4) ro-vibrational lines. (f) Modelled spectrochronogram of the convoled P(7) and O(4) ro-vibrational lines. (g) Experimental spectrochronogram of the R(5) line of the P-branch spectrum. (h) Modelled spectrochronogram of the P(6) line of the P-branch spectrum.

The coherence beating in time-resolved CRS spectra is a well-known phenomenon, due to the presence of unresolved spectral lines at the resolution of the CRS instrument. Examples of this are found in hybrid fs/ps CRS spectra of air, where some $N_2$ and $O_2$ lines cannot be resolved for probe pulse durations lesser then ~60 ps, and in high-temperature $N_2$ spectra due to the presence of unresolved hot bands[82]. The presence of significant vibrational hot bands in room-temperature CRS spectra is uncommon for diatomic molecules, with relatively high vibrational constant (e.g. ~2330 cm$^{-1}$ for $N_2$), but more complex polyatomic molecules can have vibrational modes at lower frequencies (e.g. ~667 cm$^{-1}$ for the bending mode of $CO_2$), so that excited vibrational states can have a significant Boltzmann population[54]. In order to assess the impact of vibrational hot bands on the temporal beating of the room-temperature $CH_4$ $v_2$ CRS spectrum, the time-domain model was here limited to the inclusion of fundamental transitions between the ground and first vibrationally excited states: as the model correctly reproduces the beating pattern in the experimental spectrogram, vibrational hot bands must have a negligible impact on this behaviour.

It is useful to distinguish the dynamics of the different branches in the $CH_4$ $v_2$ CRS spectrum to understand the origin of its beating. As shown in **FIG. 5(c)**, the Q-branch lines appear as a single unresolved spectral feature centred approximately at 1535 cm$^{-1}$ and shifting by ~5 cm$^{-1}$ on either side, depending on the probe pulse delay. Its beating is captured by the model in **FIG. 5(d)**, including rotational lines up to state $J$=23: it is reasonable to interpret this as the interferogram of these unresolved lines. It is then interesting to analyse the other spectral branches, which are fully resolved in **FIG. 5**. As an example, the dynamic behaviour O(4) at ~1610 cm$^{-1}$ is represented in **FIG. 5(e)** and modelled in **FIG. 5(f)**: a severe beating is still observed, which even leads to the line splitting at a probe delay of ~80 ps. While the beating can be partly attributed to the fact that the O- (and S-) branch lines are as a matter of fact not isolated, but overlap to the P- (R-) branch lines, the interference of only two harmonic contribution is expected to give rise to a simple sinusoidal pattern. A non-trivial beating pattern is also observed for line R(5) in **FIG. 5(g),(h)**, which does not overlap to any line of the S-branch and is perfectly isolated. Besides unresolved lines due to different chemical species, and to vibrational hot bands, coherence beating in time-resolved CRS spectra can arise from intra-molecular interaction, whereby the coupling of two energy degrees of freedom breaks some implicit symmetry of the molecular wave function, thus splitting otherwise degenerate energy states. This is the case e.g. in the pure-rotational CRS spectrum of $O_2$, where the spin-orbit coupling give rise to the beating of unresolved triplet transitions[83], and for the $v_2$ mode Raman spectrum of $CH_4$. In this case, the gyroscopic coupling between the ro-vibrational levels of the $v_2$ and $v_4$ modes in the $CH_4$ dyad, lifts the degeneracy of the ro-vibrational energy associated to the total angular momentum quantum number $J$. This effect can be thought of in classical terms as the introduction of a non-inertial Coriolis force[84,85] in the molecule-fixed frame of reference, which introduces an orientation-dependent contribution to the rotational energy of the molecule. The $v_2$-$v_4$ Coriolis coupling is explicitly taken into account in the tetrahedral formalism employed to compute the spectral database[86], so that the rotational sub-states associated to the same $J$ are treated as non-degenerate, and the time-domain CRS model correctly reproduce their coherence beating.

**B. Linewidth measurements**

The collisional dephasing of the $CH_4$ $v_2$ CRS spectrum is presented in **FIG. 6**.

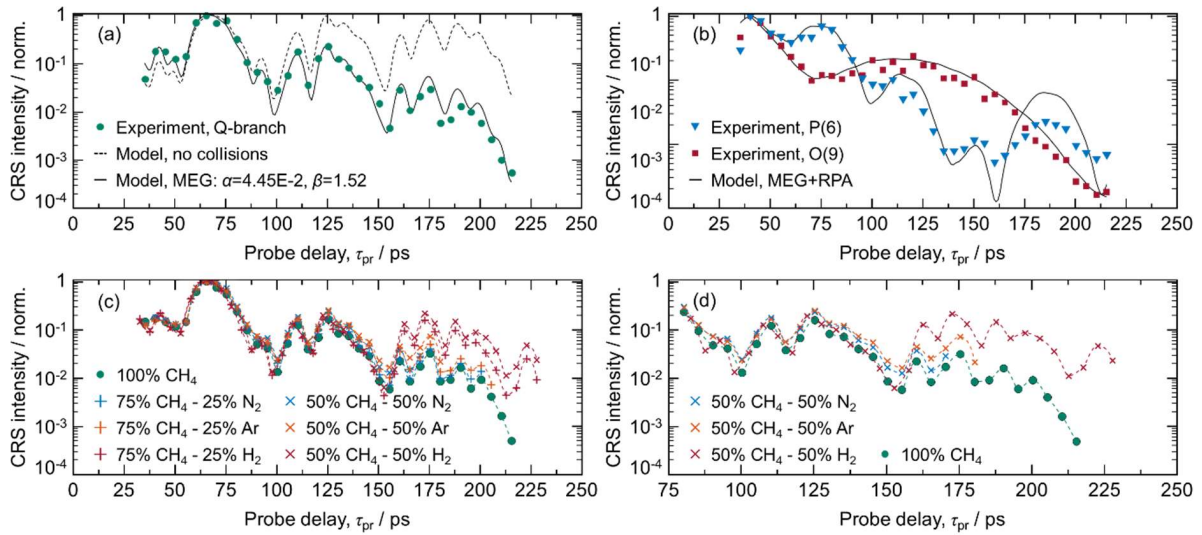

**FIG. 6.** Collisional dephasing of the $CH_4$ $v_2$ CRS signal. (a) Comparison of the experimental dephasing of the spectrally-integrated Q-branch signal and the MEG model in a room-temperature atmospheric $CH_4$ flow. (b) Experimental dephasing of isolated lines of the $CH_4$ $v_2$ spectrum in a room-temperature atmospheric $CH_4$ flow: O(9), and P(6) are selected as an example. The experimental behaviour is compared to the linewidths computed adopting the RPA. (c) Experimental dephasing of the of the spectrally-integrated isotropic Q-branch signal in binary mixtures with $N_2$, $H_2$ and argon, compared to the dephasing in a pure $CH_4$ flow. (d) Detail of the experimental dephasing of the $CH_4$ signal self-perturbed and in 50%-50% binary mixtures with $N_2$, $H_2$ and argon, for probe delays 100-225 ps.

First, the dephasing of the spectrally-integrated Q-branch signal in a room-temperature, atmospheric flow of $CH_4$ is shown in **FIG. 6(a)**: the beating of the unresolved *J*-lines in the Q-branch spectrum has a significant impact on the CRS signal intensity. The temporal behaviour of the CRS signal is thus determined by both the coherence beating and the collisional dephasing of the rotational lines. The collisional dephasing is modelled according to the MEG scaling law in Eq. (10)-(12), by fitting simultaneously the $\alpha$ and $\beta$ parameters. In order to highlight the effect of the molecular collisions, the theoretical behaviour in their absence is also given (dashed line) in **FIG. 6(a)**. The experimental dephasing of the $CH_4$ $v_2$ Q-branch spectrum and the prediction of the collisional model show a satisfactory agreement over the measured probe delay range 20-180 ps, and prove that, at least to a first approximation and at ambient conditions, the collisional dephasing of the anisotropic Q-branch lines can be well approximated by the sum rule with a MEG scaling law for the state-to-state relaxation rates. The effect of inter-branch coupling on the other hand is taken into account in the collisional dephasing model for the O-, P-, R- and S-branch lines, by introducing the RPA. The corresponding dephasing rates are thus computed according to Eq. (13) from the dephasing of the Q-branch lines, previously calculated by the MEG scaling law. **FIG. 6(b)** shows the experimental dephasing of two isolated lines in the O- and P- branches (namely O(9) and P(6)) and presents a comparison to the prediction of the RPA applied to the MEG-modelled Q-branch dephasing coefficients. The model predicts the collisional dephasing of these lines to a reasonable agreement with the experimental data, in particular for line O(6), although the predicted dephasing introduces an undue damping of the coherence beating of P(9) for probe delays larger than ~150 ps.

The effect of different collisional partners in binary mixtures with $CH_4$ was also investigated, as shown in **FIG. 6(c)**. The experimental decay of the $CH_4$ $v_2$ Q-branch signal, as measured in pure $CH_4$ or in the binary mixtures with $N_2$, $H_2$, and argon, shows a negligible dependence on the actual mixture composition for a probe delay lesser than ~80 ps. This observation confirms for the $v_2$ mode spectrum the same behaviour reported for the $CH_4$ $v_1$ Q-branch in binary mixtures with $N_2$ by Engel *et al.*[48]. Chen *et al.* defined a critical probe delay for the collisional partner independence of the $CH_4$ $v_1$ CRS signal and estimated this to be ~100 ps[52]: they experimentally confirmed it by measuring the collisional dephasing of the $CH_4$ CRS signal in binary mixtures with argon, as well as $N_2$. They employed a simple model based on gas kinetics to estimate the mean time between molecular collisions and thus define a collision sensitivity time. Below such a timescale, coherence beating dominates the temporal evolution of the $CH_4$ $v_2$ CRS signal in **FIG. 6(c)**. For probe delays larger than ~80 ps the dephasing of the signal shows sensitivity to the collisional environment, as shown more in details in **FIG. 6(d)**. $N_2$ and argon present similar behaviours as collisional partners to the $CH_4$ molecules: this and confirm the findings by Chen *et al.*, who estimated the collisional linewidths of the $CH_4$ $v_1$ Q-branch at room temperature and 500 Torr to be 0.167 cm$^{-1}$ and 0.158 cm$^{-1}$ for binary mixtures with 90% $N_2$ and argon, respectively. On the other hand, $H_2$ is a much weaker perturber to the $CH_4$ $v_2$ CRS signal as demonstrated by the comparison with the self-perturbed behaviour in **FIG. 6(d)**: at 210 ps, the $CH_4$ CRS signal in a binary mixture with 50% $H_2$ is almost two orders of magnitude larger than the corresponding signal in a flow of pure $CH_4$. This observation is explained by the largely different rotational energy manifolds of the $CH_4$ and $H_2$ molecules, such that any rotational state of the $CH_4$ scatterers has a low density of neighbouring states of the $H_2$ perturbers. The collisional RET is thus significantly lessened in case of binary mixtures of $CH_4$ and $H_2$. In the case of $N_2$, on the other hand, radiator and perturber have plenty neighbouring rotational energy states, promoting the RET in inelastic collisions.

### C. $CH_4$/air diffusion flame spectroscopy

The potential of $CH_4$ $v_2$ CRS spectroscopy for *in-situ* diagnostics in chemically reactive flows is demonstrated by performing spatially-resolved measurements across a laminar $CH_4$/air diffusion flame.

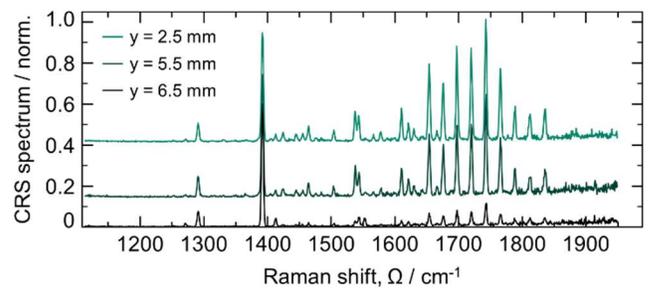

**FIG. 7.** Single-shot CRS spectra acquired at locations y=2.5, 5.5 and 6.5 mm across the $CH_4$/air diffusion flame front. The probe delay 30.7 ps is for all spectra. The peaks at 1265 and 1285 cm$^{-1}$ belong to the "red" dyad of the $CO_2$ spectrum, while the peaks at 1388 and 1409 cm$^{-1}$ belong to the "blue" dyad. $O_2$ can also be detected by its ro-vibrational Q-branch spectrum at 1553 cm$^{-1}$, which presents a clear increment moving towards the oxidiser stream. The rest of the spectral lines are attributed to the $v_2$ mode spectrum of $CH_4$.

CRS measurements over the spectral range ~1100-2000 cm$^{-1}$ were performed with a probe delay of 30.7 ps at each of the 25 locations, spaced by 0.5 mm across the flame front, moving from the fuel stream at the centre of the burner (y = 0 mm) to the oxidiser stream (y = 12 mm). The probe delay was chosen to time-gate the NR background and to maximise the signal-to-noise ratio (SNR), balancing the detection of the different branches of the $CH_4$ spectrum as visible in **FIG. 6(a)** and **FIG. 6(b)**. The single-shot detection of the $CH_4$ $v_2$ CRS signal was achieved up to location 6.5 mm, at temperatures as high as ~800 K: **FIG. 7** presents different examples of single-shot $CH_4$ $v_2$ CRS spectra acquired at three different locations (y=2.5, 5.5 and 6.5 mm) in the flame. Moving from the centre of the burner towards the chemical reaction zone of the flame, the $CH_4$ $v_2$ CRS signal is negatively impacted by the combination of increasing temperature and reducing $CH_4$ concentration so that at y=6.5 mm the SNR is reduced to less than 10. In order to extend the detection limit of the $CH_4$ $v_2$ CRS signal and validate the time-domain model at higher temperatures, 10-shot-averaged spectra were acquired at each flame location: the $CH_4$ could be detected up to 1020 K at y=8 mm. **FIG. 8** shows the ultrabroadband CRS spectra acquired across the $CH_4$/air flame front. Four different chemical species are detected in the vibrational fingerprint region of the Raman spectrum from 1100 to 2000 $cm^{-1}$: namely, the ro-vibrational $v_2$ spectrum of $CH_4$ (from 1300 to 1950 $cm^{-1}$), the $CO_2$ Fermi dyad (with fundamental Q-branch at 1285 and 1388 $cm^{-1}$), the pure-rotational spectrum of $H_2$ (four O-branch lines at 1246, 1447, 1637, and 1815 $cm^{-1}$), and the ro-vibrational spectrum of $O_2$ (with fundamental Q-branch at 1556 $cm^{-1}$). Spatially-resolved ultrabroadband CRS thus provide a window to monitor the physical-chemical processes *in-situ*, by measuring the local temperature and detecting some of the major reactants and products. The spectrum in **FIG. 8(a)** is representative of the fuel stream from the centre of the burner to y≈5 mm (as shown also in the comparison of the single-shot spectra at y=2.5 and 5.5 mm in **FIG. 7**): the spectrum is dominated by the ro-vibrational $v_2$ mode lines of $CH_4$ and by the characteristics Fermi dyad of the $CO_2$ spectrum, in particular the "blue" fundamental band of $CO_2$ at 1388 $cm^{-1}$ is the single highest spectral feature. Such an abundance of $CO_2$ at the centre of the fuel stream can be explained by buoyancy of the lightweight $CH_4$ molecule (relative molecular mass: $m$=16) against the heavier $CO_2$ ($m$=44) produced in the whole reaction zone volume of the flame. This results in the internal recirculation of the high-temperature $CO_2$ and its local mixing with the room-temperature $CH_4$ flow at the burner inlet: upon thermalisation the local temperature is higher than 296 K as attested by the clear detection of the first hot band in the $CO_2$ spectrum[54] at ~1265 and ~1409 $cm^{-1}$, and even a second hot band in the red dyad at 1244 $cm^{-1}$ as shown in the inset of **FIG. 8(a)**. It is worth noting here that this local $CO_2$ recirculation is expected to impact on the combustion chemistry due to its large heat capacity: this point will be further discussed when presenting the results of CRS thermometry. A small amount of fuel mixing with air is also visible in the spectrum: the small peak at 1556 $cm^{-1}$ is indeed the ro-vibrational Q-branch of $O_2$. This becomes much more pronounced in the high-temperature spectrum of **FIG. 8(b)**, acquired in the reaction zone at y=8.5 mm.

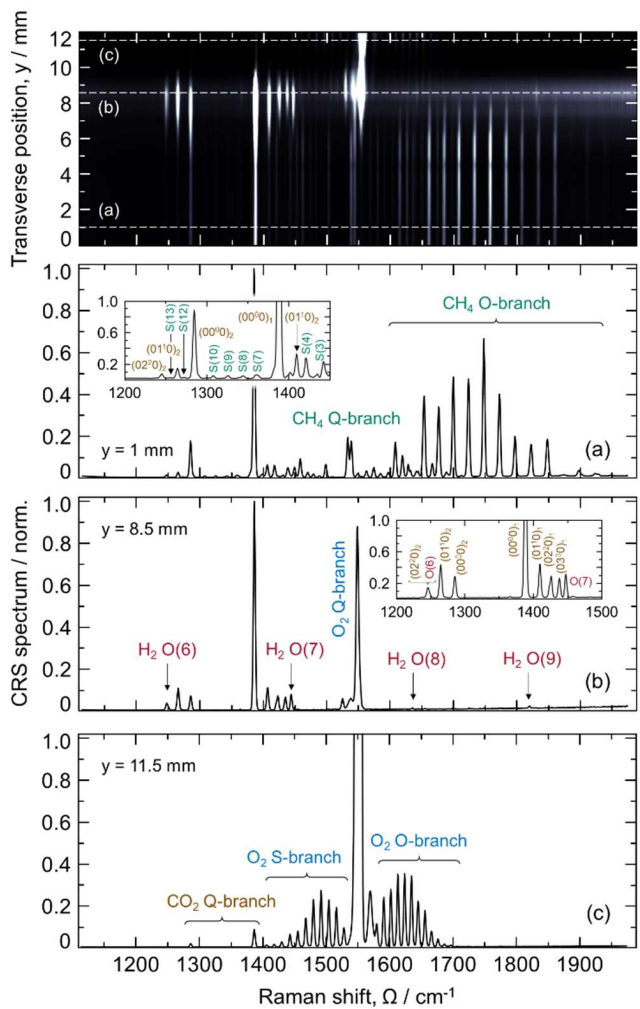

**FIG. 8.** Ultrabroadband CRS spectra in the molecular fingerprint region, measured across the laminar $CH_4$/air diffusion flame. The probe delay 30.7 ps is for all spectra. Note that gamma compression is employed in the image post-processing. (a) Fuel steam (y=1 mm): low-temperature ro-vibrational spectra of $CO_2$ and $CH_4$ (b) Reaction layer (y=8.5 mm): moving toward the reaction zone, $CH_4$ undergoes pyrolysis producing $H_2$, which is then rapidly consumed in the chemical reaction. Four lines of the pure-rotational Raman spectrum of $H_2$ are detected in the window ~1100-2000 $cm^{-1}$. The increased temperature is evident in the multiple hot bands of the ro-vibrational $CO_2$ and $O_2$ spectra (c) Oxidiser stream (y=11.5 mm): the oxidiser is ambient air, so that the Raman spectrum in the fingerprint region is dominated by the ro-vibrational spectrum of $O_2$. A small amount of the $CO_2$ produced in the combustion reaction diffuses into the oxidiser stream.

In this region of the flame the combustion reaction sustains itself by releasing heat that in turns provides the activation energy for the dissociation of the $CH_4$ molecules into radicals leading to the generation of $H_2$, which is then rapidly oxidised. Indeed the $CH_4$ $v_2$ CRS spectrum is barely detectable at this location, while four lines of the pure-rotational $H_2$ O-branch spectrum are identified: namely, O(6) at 1246 cm$^{-1}$ (overlapping with the second hot band of the $CO_2$ spectrum, as highlighted in the inset), O(7) at 1447 cm$^{-1}$, O(8) at 1637 cm$^{-1}$ and O(9) at 1815 cm$^{-1}$. The heat released in the oxidation of the fuel increases the local temperature as marked by the pronounced hot bands in the Q-branch spectra of $CO_2$, these are marked in the figure by the lowest vibrational state, denoted according to the notation in Herzberg[87] as $v_1v_2^l v_3$. As the oxidiser is ambient air, the temperature rapidly drops moving away from the reaction zone and the spectrum at y=11.5 mm is dominated by the low-temperature $O_2$ CRS signal, with the ro-vibrational O- and S-branch spectra also visible in **FIG. 8(c)**.

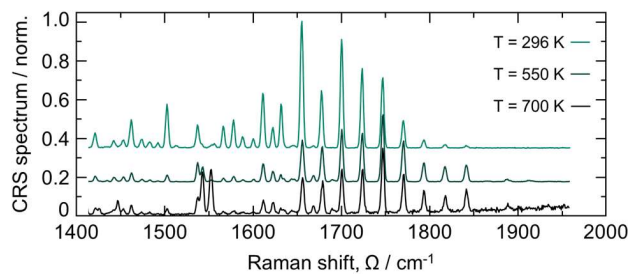

**FIG. 9.** Temperature dependence of the $CH_4$ $v_2$ CRS spectrum. Single-shot spectrum acquired in an open room-temperature $CH_4$ flow (light green), and 10-shot-averaged spectra acquired at 550 K (dark green) and 700 K (black) at location y=3.5 and 6 mm, respectively.

**FIG. 9** illustrates how the $CH_4$ $v_2$ CRS spectrum is affected by temperature, by comparing a single-shot spectrum acquired in at room temperature, and two 10-shot-averaged spectra acquired at two different locations across the flame and fitted to the time-domain model measuring the local temperature to 550 K and 700 K, respectively. As the temperature increases higher rotational state become populated, as reflected by the spread of the spectral envelope; in addition, the coherence beating of the O- and P-branch lines, and possibly of the vibrational hot bands, determines a shift of the strongest line from O(6) at room temperature to O(10) in the flame, while the probe delay is kept at 30.7 ps for all the spectra. An analogous behaviour is shown by the S-branch spectrum, with the dominating line at room temperature being S(0), and shifting to S(2) at 550 K. The small peak at 1447 cm$^{-1}$ at 700 K is attributed to line O(7) in the pure-rotational $H_2$ CRS spectrum. A significant change is also observed in the unresolved Q-branch spectrum as the temperature increases, although the effect is somewhat obscured at T=700 K because of the overlap with the ro-vibrational $O_2$ Q-branch spectrum, whose intensity increases moving towards the oxidiser stream.

Ultrabroadband CRS in the fingerprint region has so far provided us with qualitative insights into the mixing and chemical processes taking place across the diffusion flame. In order to perform quantitative measurements, it is necessary to validate the time-domain model for the $CH_4$ $v_2$ CRS spectrum at high temperature. This is done by performing direct $CH_4$ thermometry –matching the experimental 10-shot-averaged spectra up to y=6.5 mm against the synthetic library– and comparing the results to $CO_2$ CRS thermometry, developed and validated in our previous study[54]. An example of the fitting of an experimental $CH_4$ $v_2$ CRS spectrum (at y=0 mm) to the time-domain CRS model is shown in **FIG. 10(a)**. All the five ro-vibrational branches of the $CH_4$ $v_2$ spectrum in the region ~1400-1950 cm$^{-1}$ are fitted simultaneously to the synthetic spectra in the library employing a damped least-squares algorithm. As the temperature increases toward the reaction zone, the hot bands of the $CO_2$ spectrum cover the S- and R-branch lines of the $CH_4$ spectrum; similarly, the increasing oxygen concentration moving to the oxidiser stream, results in the mixing of the Q-branch lines of the $CH_4$ and $O_2$ spectra. Hence, from location y=4 mm up to 7 mm the fit was limited to the P- and O-branch lines of the $CH_4$ $v_2$ spectrum. All the five ro-vibrational branches of the $CH_4$ $v_2$ spectrum in the region ~1400-1950 cm$^{-1}$ are fitted simultaneously to the synthetic spectra in the library employing a damped least-squares algorithm. As the temperature increases toward the reaction zone, the hot bands of the $CO_2$ spectrum cover the S- and R-branch lines of the $CH_4$ spectrum; similarly, the increasing oxygen concentration moving to the oxidiser stream, results in the mixing of the Q-branch lines of the $CH_4$ and $O_2$ spectra. Hence, from location y=4 mm up to 7 mm the fit was limited to the P- and O-branch lines of the $CH_4$ $v_2$ spectrum.

**FIG. 10(b)** presents the validation of $CH_4$ $v_2$ CRS thermometry (green markers) by means of comparison with ro-vibrational $CO_2$ CRS (black markers): the abundance of $CO_2$ in the diffusion flame tested and its detection at all the measurement locations allow us to use $CO_2$ to measure the temperature profile across the whole flame front. As briefly mentioned in the previous paragraph, the large concentration of $CO_2$ in the centre of the burner, as detected in the ultrabroadband CRS spectra of **FIG. 8**, has a significant impact on the combustion chemistry, given the large heat capacity of the $CO_2$ molecules, which thus act as heat sinks in the reaction zone on the flame. The result is a smoother temperature profile than expected for a laminar CH$_4$/air diffusion flame[88]: the temperature at the centre of the fuel stream (y=0 mm) is larger than the room-temperature of the inlet CH$_4$ flow, while the maximum temperature in the reaction zone (~1430 K) is significantly lower than expected for a laminar axisymmetric methane/air diffusion flame[88]. The thermal effect of the back-diffusing CO$_2$ on the measured flame temperature in our experiment is comparable to the reduction in the adiabatic flame temperature measured in laminar premixed CH$_4$/air flames for a 20% CO$_2$ dilution[89]. Comparing the temperature measurements obtained by CH$_4$ $\nu_2$ CRS and CO$_2$ CRS we can validate our time-domain CH$_4$ CRS model at temperatures as high as ~800 K, and quantify the accuracy of CH$_4$ $\nu_2$ CRS thermometry. The two thermometric techniques show a very satisfactory agreement not only in terms of the average temperature measured at each flame location, but also within each temporal sequence of 1000 frames, as shown in **FIG. 10(c)** for y=0 mm.

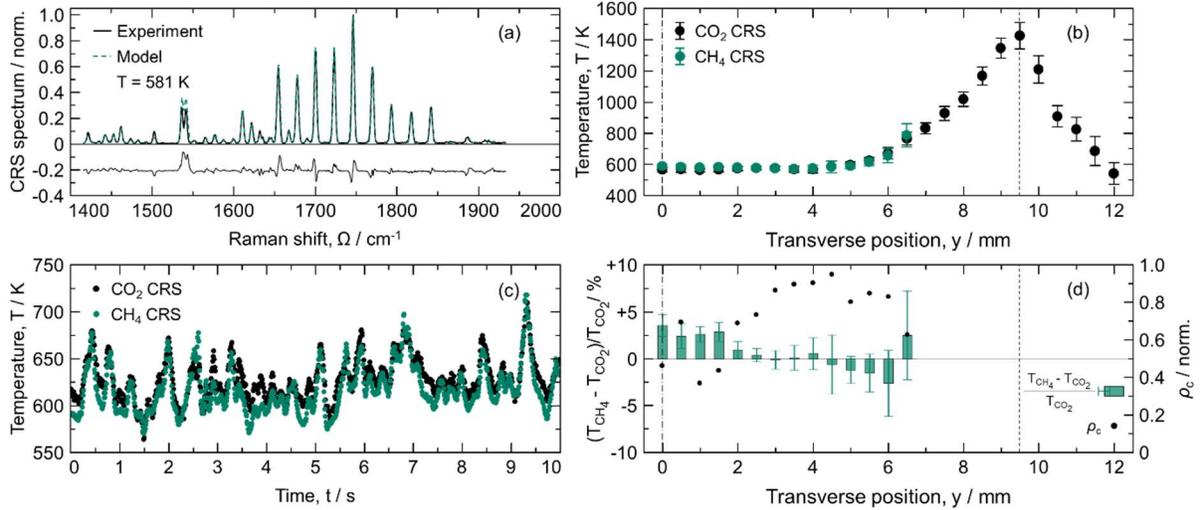

**FIG. 10.** Ro-vibrational CH$_4$ $\nu_2$ CRS thermometry. (a) Experimental 10-shot-averaged CH$_4$ $\nu_2$ CRS spectrum acquired at location y=0 mm and comparison to the time-domain CRS model for thermometry. (b) Average temperature profile across the laminar CH$_4$/air diffusion flame as measured by ultrabroadband fs/ps CRS thermometry: ro-vibrational CO$_2$ thermometry (black) provides validation to ro-vibrational CH$_4$ thermometry (green). (c) Comparison of CO$_2$ and CH$_4$ CRS thermometry over sample of 1000 ultrabroadband spectra, acquired at y=5.5 mm: the same dynamics is reproduced by the two independent methods, which proves the physical nature of the temperature oscillations. (d) Accuracy and precision of CH$_4$ $\nu_2$ CRS thermometry as compared to CO$_2$ CRS thermometry: the thermometric accuracy (green bars) is better than 3% at all measurement locations, while the precision is fundamentally limited by the physical fluctuations in the flame. Concordance correlation (black dots) quantifies the agreement of the temperature dynamics measured by CO$_2$ and CH$_4$ CRS at each location.

The agreement between CH$_4$ $\nu_2$ and CO$_2$ CRS thermometry is quantified in **FIG. 10(d)** evaluating the systematic bias between the two methodologies (bar plot), and by means of Lin's concordance correlation coefficient, defined as[90]:

$$\rho_c = \frac{2\sigma_{CO_2,CH_4}}{\sigma_{CO_2}^2 + \sigma_{CH_4}^2 + \left(\mu_{CO_2} - \mu_{CH_4}\right)^2} \quad (14)$$

where $\mu_{CO_2}$ and $\mu_{CH_4}$ are the mean temperature measured by CO$_2$ and CH$_4$ CRS respectively, with corresponding standard deviations $\sigma_{CO_2}$, $\sigma_{CO_4}$, and covariance $\sigma_{CO_2,CH_4}$.

The formula in Eq. (14) quantifies the correlation between the two temperature measurements independently of the possible systematic bias between them and of the temperature fluctuations in the sample. The temperature dynamics measured by CO$_2$ CRS thermometry can thus be adequately reproduced by CH$_4$ $\nu_2$ CRS, with an accuracy better than 3% at all measurements location. The satisfactory frame-by-frame comparison between CO$_2$ and CH$_4$ $\nu_2$ CRS thermometry, as quantifies by the concordance correlation factor, also allow us two establish the origin of the fluctuations in the measured temperature, due to oscillations in the flame.

**FIG. 11(a)** shows the spectra of the temperature dynamics as measured by CO$_2$ and CH$_4$ $\nu_2$ CRS thermometry at y=3 mm and represented by circles in **FIG. 11(b)**. These spectra present common frequencies in the range ~0.1-12.5 Hz (the upper limit being the value where the spectra of $T_{CH_4}$ and $T_{CO_2}$ diverge), corresponding to the frequencies associated to the flame oscillations. Applying a spectral filter in this window[91], it is possible to isolate the inherent noise in the measurements, represented by the solid lines in **FIG. 11(b)**, even though a lower-frequency oscillation (<0.1 Hz), unresolved over the 10 s acquisition window, can still be made out in the smoothened profiles. The thermometric precision was thus improved from 1.9% for CH$_4$ $\nu_2$ CRS and 2.4 % for CO$_2$ CRS to 0.57% and 0.91%, respectively. The identification of the frequency at which the two spectra diverge depends on the metric used to define this divergence, but this has only a minor effect on the quantification of the measurement precision: filtering oscillations only up to 10 Hz resulted in a precision of 0.61% for $CH_4$ $v_2$ CRS and 1.0% for $CO_2$ CRS, while extending the filter to 15 Hz changed it to 0.54% and 0.90% respectively. The 12.5 Hz value was found to be the lowest applicable to all the datasets acquired in the flame: the resulting thermometric precision was better than 2% at all measurements locations for both $CO_2$ and $CH_4$ $v_2$ CRS thermometry.

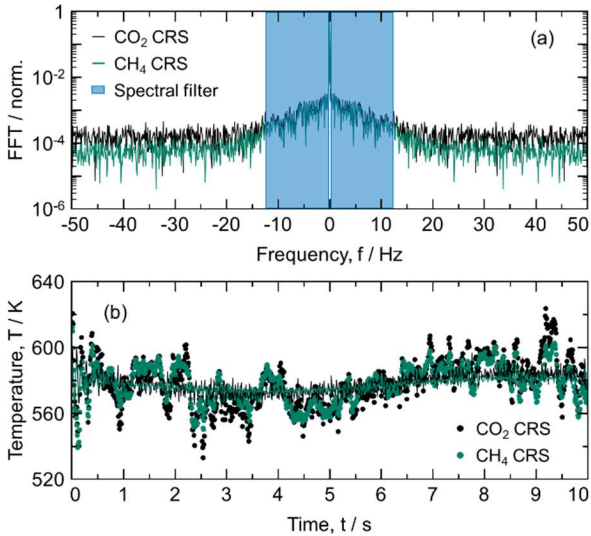

**FIG. 11.** Temperature dynamics measured by ro-vibrational CRS thermometry performed on $CO_2$ (black) and $CH_4$ (green). (a) Spectrum of the temperature dynamics: common frequencies, due to physical oscillations in the flame environment, are identified in the range ~0.1-12.5 Hz. (b) Original (circle) and filtered (solid line) temperature dynamics: removing the temperature fluctuations with common frequencies we estimate the inherent precision of $CH_4$ $v_2$ CRS thermometry to be better than 1% for this sample.

## IV. CONCLUSIONS

We have reported the first investigation of the $v_2$ mode Raman spectrum of $CH_4$ by means of coherent Raman spectroscopy: ultrabroadband two-beam fs/ps CRS was employed to perform time-resolved measurements of the ro-vibrational $CH_4$ $v_2$ spectrum and to demonstrate its application as a combustion diagnostic tool. Our CRS instrument employs a single regenerative fs amplifier to generate both broadband fs and narrowband ps pulses; fs laser-induced filamentation is employed *in-situ* to compress the fs pulse to <20 fs, so as to excite the ro-vibrational Raman modes in the vibrational fingerprint region. This spectral region is of particular interest in the investigation of chemical reactions in gas-phase environments, as a number of Raman-active species have a spectral signature in the range ~1100-2000 $cm^{-1}$.

We developed a time-domain CRS model for the $CH_4$ $v_2$ spectrum employing the MeCaSDa calculated spectroscopic database with the position and cross section of ~10 million Raman lines, as computed at the University of Burgundy[72,73]. We employed this line list by rescaling the spontaneous Raman cross section by the differential Boltzmann population between the two ro-vibrational states involved in the CRS process, and we applied a spectral filter, considering only CRS lines stronger than a cut-off value of the strongest lines, to reduce the computational time of the temperature-dependent synthetic CRS libraries. We then validated the CRS model by performing $CH_4$ $v_2$ thermometry in a laminar $CH_4$/air diffusion flame: the temperature estimations by $CH_4$ $v_2$ CRS were found to converge for cut-off values smaller than 1E-6, for all the measured temperatures up to ~800 K.

The CRS model include collisional Raman linewidths computed by a modified energy-gap scaling law for the Q-branch lines and adopting the random phase approximation for the derivation of the O-, P-, R- and S-branch collisional dephasing rates. The collisional dephasing of the self-perturbed CRS signal in a room-temperature $CH_4$ flow is used to fit the $\alpha$ and $\beta$ parameters of the MEG model. The use of the sum rule for the calculation of the total collisional dephasing rate of the anisotropic Q-branch for the completely depolarized $CH_4$ $v_2$ Raman spectrum is a rather coarse approximation, but it is here justified by the fact that only atmospheric measurements are reported. The extension of the present model to high pressure measurements should include the impact of inelastic collisions, determining molecular reorientation[92,93], and the effect of intra-branch coupling of the vibrationally-degenerate lines[94] of the $CH_4$ $v_2$ spectrum. The effect of different collisional partners was furthermore investigated by performing dephasing measurements in binary mixtures of $CH_4$ with $N_2$, $H_2$, and Ar, in varying concentrations. We employed the model to study the time-domain behaviour of the $CH_4$ $v_2$ CRS spectrum at room temperature: a strong coherence beating is observed, not only for the unresolved rotational lines in the Q-branch, but also for the well-resolved lines in the other branches. The beating of perfectly isolated in the P- and R-branch spectra, in particular, was demonstrated to be due to the Coriolis splitting[23,84,85] of the fine structure of the rotational states due to the different symmetry of the wave function components, and was well captured by our the time-domain CRS model.

We then performed spatially resolved ultrabroadband fs/ps CRS measurements across the laminar $CH_4$/air diffusion flame front, identifying the Raman signature of four major combustion species. We demonstrated the single-shot detection of the $CH_4$ $\nu_2$ CRS signal at temperatures as high as ~800 K, and we employed the ro-vibrational $CO_2$ CRS spectrum, detected at all measurement locations, to perform (10-shot-averaged) CRS thermometry across the flame front and to validate $CH_4$ $\nu_2$ thermometry. The accuracy of $CH_4$ $\nu_2$ thermometry was better than 2% at all measurement locations, while the precision was intrinsically limited by fluctuations in the flame, as proven by the same temperature dynamics being measured independently by $CO_2$ and $CH_4$ $\nu_2$ CRS.

The detection of $CH_4$, $CO_2$, $O_2$ and $H_2$ across the flame front is particularly appealing for the prospect use of ultrabroadband fs/ps CRS the *in-situ* investigation of $CH_4$ pyrolysis and chemical reforming in non-equilibrium environments, such as plasma reactors, for e.g. production of turquoise hydrogen[95]. The extension of the present work to concentration measurements is currently limited by the lack of reported data on the absolute Raman cross section for the $CH_4$ $\nu_2$ mode, which are required to meaningfully compare the CRS spectra of different chemical species, and should be addressed in future works. An additional difficulty is represented by the unknown spectral excitation efficiency provided by the compressed supercontinuum generated by filamentation in the flame environment: in a recent work we have demonstrated a novel experimental protocol for the *in-situ* referencing of the ultrabroadband spectral excitation[57]. The implementation of this protocol is based on the polarisation control of the CRS signal generation and required the Raman spectrum to have a depolarisation ratio larger than 0.5. The completely depolarised $CH_4$ $\nu_2$ CRS signal could thus be generated with the same polarisation as the pure-rotational $H_2$ CRS signal for accurate $H_2$/$CH_4$ concentration measurements via ultrabroadband fs/ps CRS.

Furthermore, the modelling of the $CH_4$ $\nu_2$ Raman spectrum can serve as blueprint for heavier hydrocarbon molecules, such as ethane and dimethyl ether, which also have a Raman-active vibrational mode due to the bending of the H-C-H bond[27,96]. The availability of spectral data for such molecules could pave the way to the extension of ultrabroadband fs/ps CRS to the *in-situ* investigation of oxy-fuel combustion in a number of practical applications[97].

## ACKNOWLEDGEMENTS

We acknowledge the financial support provided by the Netherlands Organization for Scientific Research (NWO), obtained through a Vidi grant in the Applied and Engineering Sciences domain (AES) (15690). A. Bohlin is thankful for support through the RIT (Space for Innovation and Growth) project/European Regional Development Fond in Kiruna, Sweden.

## REFERENCES

[1] J.I. Lunine and S.K. Atreya, Nat. Geosci. **1**, 159 (2008).

[2] C.R. Webster, P.R. Mahaffy, S.K. Atreya, J.E. Moores, G.J. Flesch, C. Malespin, C.P. Mckay, G. Martinez, C.L. Smith, J. Martin-Torres, J. Gomez-Elvira, M.-P. Zorzano, M.H. Wong, M.G. Trainer, A. Steele, D. Archer Jr., B. Sutter, P.J. Coll, C. Freissinet, P.-Y. Meslin, R. V. Gough, C.H. House, A. Pavlov, J.L. Eigenbrode, D.P. Glavin, J.C. Pearson, D. Keymeulen, L.E. Christensen, S.P. Schwenzer, R. Navarro-Gonzalez, J. Pla-García, S.C.R. Rafkin, Á. Vicente-Retortillo, H. Kahanpää, D. Viudez-Moreiras, M.D. Smith, A.-M. Harri, M. Genzer, D.M. Hassler, M. Lemmon, J. Crisp, S.P. Sander, R.W. Zurek, and A.R. Vasavada, Science **360**, 1093 (2018).

[3] L. Ernst, B. Steinfeld, U. Barayeu, T. Klintzsch, M. Kurth, D. Grimm, T.P. Dick, J.G. Rebelein, I.B. Bischofs, and F. Keppler, Nature **603**, 482 (2022).

[4] V. Masson-Delmotte, P. Zhai, A. Pirani, S.L. Connors, C. Péan, S. Berger, N. Caud, Y. Chen, L. Goldfarb, M.I. Gomis, M. Huang, K. Leitzell, E. Lonnoy, J.B.R. Matthews, T.K. Maycock, T. Waterfield, O. Yelekçi, R. Yu, and B. Zhou, editors, *IPCC, 2021: Climate Change 2021: The Physical Science Basis. Contribution of Working Group I to the Sixth Assessment Report of the Intergovernmental Panel on Climate Change* (Cambridge University Press, Cambridge, United Kingdom and New York, NY, USA, 2021).

[5] T.W. Hesterberg, C.A. Lapin, and W.B. Bunn, Environ. Sci. Technol. **42**, 6437 (2008).

[6] D.E. Holmes and J.A. Smith, Adv. Appl. Microbiol. **97**, 1 (2016).

[7] M. Qi, J. Lee, S. Hong, J. Kim, Y. Liu, J. Park, and I. Moon, Energy **256**, 124583 (2022).

[8] L. Bromberg, D.R. Cohn, A. Rabinovich, C. O'Brien, and S. Hochgreb, Energy and Fuels **12**, 11 (1998).

[9] J.R. Fincke, R.P. Anderson, T. Hyde, B.A. Detering, R. Wright, R.L. Bewley, D.C. Haggard, and W.D. Swank, Plasma Chem. Plasma Process. **22**, 105 (2002).

[10] A.I. Olivos-Suarez, À. Szécsényi, E.J.M. Hensen, J. Ruiz-Martinez, E.A. Pidko, and J. Gascon, ACS Catal. **6**, 2965 (2016).

[11] S.-I. Chou, D.S. Baer, and R.K. Hanson, Appl. Opt. **36**, 3288 (1997).

[12] P. Werle and A. Popov, Appl. Opt. **38**, 1494 (1999).

[13] A.A. Kosterev, R.F. Curl, F.K. Tittel, C. Gmachl, F. Capasso, D.L. Sivco, J.N. Baillargeon, A.L. Hutchinson, and A.Y. Cho, Opt. Lett. **24**, 1762 (1999).

[14] E. Baumann, F.R. Giorgetta, W.C. Swann, A.M. Zolot, I. Coddington, and N.R. Newbury, Phys. Rev. A **84**, (2011).

[15] P. Werle, Spectrochim. Acta Part A **54**, 197 (1998).

[16] C.S. Goldenstein, R.M. Spearrin, J.B. Jeffries, and R.K.

Hanson, Prog. Energy Combust. Sci. **60**, 132 (2017).

[17] I.E. Gordon, L.S. Rothman, C. Hill, R. V Kochanov, Y. Tan, P.F. Bernath, M. Birk, V. Boudon, A. Campargue, K. V Chance, B.J. Drouin, J. Flaud, R.R. Gamache, J.T. Hodges, D. Jacquemart, V.I. Perevalov, A. Perrin, K.P. Shine, M.H. Smith, J. Tennyson, G.C. Toon, H. Tran, V.G. Tyuterev, A. Barbe, A.G. Császár, V.M. Devi, T. Furtenbacher, J.J. Harrison, J. Hartmann, A. Jolly, T.J. Johnson, T. Karman, I. Kleiner, A.A. Kyuberis, J. Loos, O.M. Lyulin, S.T. Massie, S.N. Mikhailenko, N. Moazzen-ahmadi, H.S.P. Müller, O. V Naumenko, A. V Nikitin, O.L. Polyansky, J. Vander Auwera, G. Wagner, J. Wilzewski, P. Wcis, S. Yu, and E.J. Zak, J. Quant. Spectrosc. Radiat. Transf. **203**, 3 (2017).

[18] R.J. Tancin and C.S. Goldenstein, Opt. Express **29**, 30140 (2021).

[19] W. Cai and C.F. Kaminski, Prog. Energy Combust. Sci. **59**, 1 (2017).

[20] C. Wei, K.K. Schwarm, D.I. Pineda, and R.M. Spearrin, Opt. Lett. **45**, 2447 (2020).

[21] T.D. Butterworth, B. Amyay, D. v. d. Bekerom, A. v. d. Steeg, T. Minea, N. Gatti, Q. Ong, C. Richard, C. van Kruijsdijk, J.T. Smits, A.P. van Bavel, V. Boudon, and G.J. van Rooij, J. Quant. Spectrosc. Radiat. Transf. **236**, 106562 (2019).

[22] T.F. Guiberti, Y. Krishna, W.R. Boyette, C. Yang, W.L. Roberts, and G. Magnotti, Proc. Combust. Inst. **38**, 1647 (2021).

[23] T. Feldman, J. Romanko, and H.L. Welsh, Can. J. Phys. **33**, 138 (1955).

[24] M.A. Thomas and H.L. Welsh, Can. J. Phys. **38**, 1291 (1960).

[25] W. Meier, R.S. Barlow, Y.L. Chen, and J.Y. Chen, Combust. Flame **123**, 326 (2000).

[26] R.S. Barlow, S. Meares, G. Magnotti, H. Cutcher, and A.R. Masri, Combust. Flame **162**, 3516 (2015).

[27] G. Magnotti, U. KC, P.L. Varghese, and R.S. Barlow, J. Quant. Spectrosc. Radiat. Transf. **163**, 80 (2015).

[28] D. Butz, S. Hartl, S. Popp, S. Walther, R.S. Barlow, C. Hasse, A. Dreizler, and D. Geyer, Combust. Flame **210**, 426 (2019).

[29] D. V. Petrov, I.I. Matrosov, A.R. Zaripov, and A.S. Maznoy, Appl. Spectrosc. **74**, 948 (2020).

[30] T. Butterworth, A. van de Steeg, D. van den Bekerom, T. Minea, T. Righart, Q. Ong, and G. van Rooij, Plasma Sources Sci. Technol. **29**, 095007 (2020).

[31] R.R. Smith, D.R. Killelea, D.F. DelSesto, and A.L. Utz, Science **304**, 992 (2004).

[32] J.W.L. Lee, D.S. Tikhonov, P. Chopra, S. Maclot, A.L. Steber, S. Gruet, F. Allum, R. Boll, X. Cheng, S. Düsterer, B. Erk, D. Garg, L. He, D. Heathcote, M. Johny, M.M. Kazemi, H. Köckert, J. Lahl, A.K. Lemmens, D. Loru, R. Mason, E. Müller, T. Mullins, P. Olshin, C. Passow, J. Peschel, D. Ramm, D. Rompotis, N. Schirmel, S. Trippel, J. Wiese, F. Ziaee, S. Bari, M. Burt, J. Küpper, A.M. Rijs, D. Rolles, S. Techert, P. Eng-Johnsson, M. Brouard, C. Vallance, B. Manschwetus, and M. Schnell, Nat. Commun. **12**, 6107 (2021).

[33] C.A. Marx, U. Harbola, and S. Mukamel, Phys. Rev. A **77**, 1 (2008).

[34] G. Millot, B. Lavorel, R. Chaux, G. Pierre, H. Berger, J.I. Steinfeld, and B. Foy, J. Mol. Spectrosc. **127**, 156 (1988).

[35] D. Bermejo, J. Santos, and P. Cancio, J. Mol. Spectrosc. **156**, 15 (1992).

[36] J.J. Barrett and R.F. Begley, Appl. Phys. Lett. **27**, 129 (1975).

[37] D.N. Kozlov and V. V. Smirnov, Pis'ma Zh. Eksp. Teor. Fiz. **26**, 31 (1977).

[38] D.N. Kozlov, A.M. Prokhorov, and V. V. Smirnov, J. Mol. Spectrosc. **77**, 21 (1979).

[39] H. Frunder, D. Illig, H. Finsterhölzl, H.W. Schrotter, B. Lavorel, G. Roussel, J.C. Hilico, J.P. Champion, G. Pierre, G. Poussigue, and E. Pascaud, Chem. Phys. Lett. **100**, 110 (1983).

[40] G. Millot, B. Lavorel, and J.I. Steinfeld, J. Chem. Phys. **95**, 7938 (1991).

[41] M. Ridder, A.A. Suvernev, and T. Dreier, J. Chem. Phys. **105**, 3376 (1996).

[42] M.L. Strekalov, Mol. Phys. **100**, 1049 (2002).

[43] T. Dreier, B. Lange, J. Wolfrum, and M. Zahn, Appl. Phys. B **45**, 183 (1988).

[44] E. Jourdanneau, T. Gabard, F. Chaussard, R. Saint-Loup, H. Berger, E. Bertseva, and F. Grisch, J. Mol. Spectrosc. **246**, 167 (2007).

[45] F. Grisch, E. Bertseva, M. Habiballah, E. Jourdanneau, F. Chaussard, R. Saint-Loup, T. Gabard, and H. Berger, Aerosp. Sci. Technol. **11**, 48 (2007).

[46] R. Leonhardt, W. Holzapfel, W. Zinth, and W. Kaiser, Chem. Phys. Lett. **133**, 373 (1987).

[47] B.D. Prince, A. Chakraborty, B.M. Prince, and H.U. Stauffer, J. Chem. Phys. **125**, 044502 (2006).

[48] S.R. Engel, J.D. Miller, C.E. Dedic, T. Seeger, A. Leipertz, and T.R. Meyer, J. Raman Spectrosc. **34**, 1336 (2013).

[49] C.N. Dennis, D.L. Cruise, H. Mongia, G.B. King, and R.P. Lucht, in *53rd AIAA Aerosp. Sci. Meet.* (2015), pp. 1–13.

[50] A. Bohlin and C.J. Kliewer, Appl. Phys. Lett. **104**, 031107 (2014).

[51] A. Bohlin, C. Jainski, B.D. Patterson, A. Dreizler, and C.J. Kliewer, Proc. Combust. Inst. **36**, 4557 (2017).

[52] T.Y. Chen, C.J. Kliewer, B.M. Goldberg, E. Kolemen, and Y. Ju, Combust. Flame **224**, 183 (2021).

[53] Y. Ran, A. Boden, F. Küster, F. An, A. Richter, S. Guhl, S. Nolte, and R. Ackermann, Appl. Phys. Lett. **119**, 243905 (2021).

[54] F. Mazza, N. Griffioen, L. Castellanos, D. Kliukin, and A. Bohlin, Combust. Flame **237**, 111738 (2022).

[55] L. Castellanos, F. Mazza, D. Kliukin, and A. Bohlin, Opt. Lett. **45**, 4662 (2020).

[56] Z. Tian, H. Zhao, Y. Gao, H. Wei, Y. Tan, and Y. Li, Appl. Phys. Lett. **121**, 081102 (2022).

[57] F. Mazza, A. Stutvoet, L. Castellanos, D. Kliukin, and A. Bohlin, Opt. Express **30**, 35232 (2022).

[58] F. Träger, *Springer Handbook of Laser and Optics* (Springer, New York, 2007).
[59] A. Bohlin, B.D. Patterson, and C.J. Kliewer, J. Chem. Phys. **138**, 081102 (2013).
[60] J.H. Odhner, D.A. Romanov, and R.J. Levis, Phys. Rev. Lett. **103**, 075005 (2009).
[61] J.D. Miller, S. Roy, M.N. Slipchenko, J.R. Gord, and T. Meyer, Opt. Express **19**, 15627 (2011).
[62] S.P. Kearney, D.J. Scoglietti, and C.J. Kliewer, Opt. Express **21**, 12327 (2013).
[63] T.L. Courtney, A. Bohlin, B.D. Patterson, and C.J. Kliewer, J. Chem. Phys. **146**, 224202 (2017).
[64] J.D. Miller, M.N. Slipchenko, T.R. Meyer, H.U. Stauffer, and J.R. Gord, Opt. Lett. **35**, 2430 (2010).
[65] N. Owschimikow, F. Königsmann, J. Maurer, P. Giese, A. Ott, B. Schmidt, and N. Schwentner, J. Chem. Phys. **133**, 044311 (2010).
[66] F. Metz, W.E. Howard, L. Wunsch, H.J. Neusser, and E.W. Schlag, Proc. R. Soc. Lond. A **363**, 381 (1978).
[67] A.H. Nielsen and H.H. Nielsen, Phys. Rev. **48**, 864 (1935).
[68] M.J. Hollas, *High Resolution Spectroscopy*, 1st ed. (Butterworths, London, 1982).
[69] E. Bright Wilson, J. Chem. Phys. **3**, 276 (1935).
[70] V. Boudon, J. Champion, T. Gabard, M. Lo, M. Rotger, and C. Wenger, in *Handb. High-Resolution Spectrosc.*, edited by M. Quack and F. Merkt (John Wiley & Sons, Ltd., 2011).
[71] J.-P. Champion, M. Loéte, and G. Pierre, *Spherical Top Spectra* (1992).
[72] Y.A. Ba, C. Wenger, R. Surleau, V. Boudon, M. Rotger, L. Daumont, D.A. Bonhommeau, V.G. Tyuterev, and M.L. Dubernet, J. Quant. Spectrosc. Radiat. Transf. **130**, 62 (2013).
[73] C. Richard, V. Boudon, and M. Rotger, J. Quant. Spectrosc. Radiat. Transf. **251**, 107096 (2020).
[74] B. Amyay and V. Boudon, J. Quant. Spectrosc. Radiat. Transf. **219**, 85 (2018).
[75] A.S. Tanichev and D. Petrov, J. Raman Spectrosc. **53**, 654 (2021).
[76] E. Jourdanneau, F. Chaussard, R. Saint-loup, T. Gabard, and H. Berger, J. Mol. Spectrosc. **233**, 219 (2005).
[77] L.A. Rahn and R.E. Palmer, J. Opt. Soc. Am. B **3**, 1164 (1986).
[78] T. Seeger, F. Beyrau, A. Braeuer, and A. Leipertz, J. Raman Spectrosc. **34**, 932 (2003).
[79] G. Fanjoux, B. Lavorel, and G. Millot, J. Raman Spectrosc. **29**, 391 (1998).
[80] J. Boissoles, C. Boulet, L. Bonamy, and D. Robert, J. Quant. Spectrosc. Radiat. Transf. **42**, 509 (1989).
[81] A.E. DePristo, S.D. Augustin, R. Ramaswamy, and H. Rabitz, J. Chem. Phys. **71**, 850 (1979).
[82] T.Y. Chen, N. Liu, C.J. Kliewer, A. Dogariu, E. Kolemen, and Y. Ju, Opt. Lett. **47**, 1351 (2022).
[83] T.L. Courtney and C.J. Kliewer, J. Chem. Phys. **149**, 234201 (2018).
[84] H.A. Jahn, Proc. R. Soc. London. Ser. A Math. Phys. Sci. **168**, 469 (1938).
[85] H.A. Jahn, Phys. Rev. **56**, 680 (1939).
[86] J.P. Champion and G. Pierre, J. Mol. Spectrosc. **79**, 255 (1980).
[87] G. Herzberg, *Molecular Spectra and Molecular Structure. II. Infrared and Raman Spectra of Polyatomic Molecules*, 10th ed. (Van Nostrand Reinhold Company, New York, 1945).
[88] M.D. Smooke, P. Lin, J.K. Lam, and M.B. Long, Proc. Combust. Inst. **23**, 575 (1991).
[89] C. Zhang, G. Hu, S. Liao, Q. Cheng, C. Xiang, and C. Yuan, Energy **106**, 431 (2016).
[90] L.I. Lin, Biometrics **45**, 255 (1989).
[91] J. Barros, M. Scherman, E. Lin, N. Fdida, R. Santagata, B. Attal-Tretout, and A. Bresson, Opt. Express **28**, 34656 (2020).
[92] S.I. Temkin, L. Bonamy, J. Bonamy, and D. Robert, Phys. Rev. A **47**, 1543 (1993).
[93] L. Bonamy, J. Bonamy, D. Robert, S.I. Temkin, G. Millot, and B. Lavorel, J. Chem. Phys. **101**, 7350 (1994).
[94] A.S. Pine, J. Quant. Spectrosc. Radiat. Transf. **57**, 145 (1997).
[95] N.N. Morgan and M. ElSabbagh, Plasma Chem. Plasma Process. **37**, 1375 (2017).
[96] Ö. Andersson, H. Neij, J. Bood, B. Axelsson, and M. Aldén, Combust. Sci. Technol. **137**, 299 (1998).
[97] J.W. Tröger, C. Meißner, and T. Seeger, J. Raman Spectrosc. **47**, 1149 (2016).